\pdfoutput=1
\documentclass[reprint,aps,prb,floatfix,onecolumn,superscriptaddress,showpacs,
amsmath,amssymb,showpacs,12pt]{revtex4-1}

\usepackage{graphicx}
\usepackage{epstopdf}
\usepackage{epsfig}
\usepackage{epsf}
\usepackage{url}
\usepackage[USenglish]{babel}
\usepackage{hyperref}
\def\bcen{\begin{center}}
\def\ecen{\end{center}}
\allowdisplaybreaks
\renewcommand\[{\begin{equation}}
\renewcommand\]{\end{equation}}
\usepackage{verbatim}
\usepackage{xcolor}
\usepackage{natbib}
\usepackage{dblfloatfix}
\usepackage{amsmath, nccmath}
\usepackage[export]{adjustbox}
\usepackage{dcolumn}
\usepackage{bm}
\usepackage{bbm}
\usepackage{lipsum}
\usepackage{tikz-feynman} 
\usepackage{float}
\usepackage{mleftright}
\usepackage{verbatim}
\begin{document}
\title{Normal state of Nd$_{1-x}$Sr$_x$NiO$_2$ from self-consistent $GW$+EDMFT}
\author{Francesco Petocchi}
\affiliation{Department of Physics, University of Fribourg, 1700 Fribourg, 
Switzerland}
\author{Viktor Christiansson}
\affiliation{Department of Physics, University of Fribourg, 1700 Fribourg, 
Switzerland}
\author{Fredrik Nilsson}
\affiliation{Department of Physics, Division of Mathematical Physics, Lund 
University, Professorsgatan 1, 223 63 Lund, Sweden}
\author{Ferdi Aryasetiawan}
\affiliation{Department of Physics, Division of Mathematical Physics, Lund 
University, Professorsgatan 1, 223 63 Lund, Sweden}
\author{Philipp Werner}
\affiliation{Department of Physics, University of Fribourg, 1700 Fribourg, 
Switzerland}
\begin{abstract}
The recent discovery of superconductivity in hole-doped NdNiO$_2$ thin films has captivated the condensed matter physics community. Such compounds with a formal Ni$^+$ valence have been theoretically proposed as possible analogues of the cuprates, and the exploration of their electronic structure and pairing mechanism may provide important insights into the phenomenon of unconventional superconductivity. At the modeling level, there are however fundamental issues that need to be resolved. While it is generally agreed that the low-energy properties of cuprates can to a large extent be captured by a single-band model, there has been a controversy in the recent literature about the importance of a multi-band description of the nickelates. The origin of this controversy is that studies based entirely on density functional theory (DFT) calculations miss important correlation and multi-orbital effects induced by Hund coupling, while model calculations or simulations based on the combination of DFT and (extended) dynamical mean field theory ((E)DMFT) involve ad-hoc parameters and double counting corrections that substantially affect the results. Here we use a multi-site extension of the recently developed $GW$+EDMFT method, which is free of adjustable parameters, to self-consistently compute the interaction parameters and electronic structure of hole-doped NdNiO$_2$. This full ab-initio simulation demonstrates the importance of a multi-orbital description, even for the undoped compound, and produces results for the resistivity and Hall conductance in qualitative agreement with experiment.  
\end{abstract}

\maketitle
%
%
%
%
 
NdNiO$_2$ is isostructural with CaCuO$_2$, which exhibits high-temperature superconductivity upon hole doping. Ni is coordinated with two oxygens, forming NiO$_2$ square planes, separated by Nd as a spacer cation. \cite{Anisimov1999,Lee2004,Nomura2019} It has the uncommon oxidation value of +1, resulting in a $d^9$ electronic configuration. Despite these obvious similarities with the  cuprate superconductors, there are also relevant differences between the nickelates and cuprates. The energy splitting between the Ni 3$d$ levels and the O 2$p$ levels is almost twice larger than the corresponding splitting  in CaCuO$_2$,\cite{Lee2004,Jiang2019,Lechermann2020a} which puts NdNiO$_2$ into the Hubbard regime rather than the charge transfer regime,\cite{Zaanen1985} and possibly precludes the formation of Zhang-Rice singlets. The peculiar structure with missing apical oxygens significantly alters the crystal field: several DFT calculations\cite{Sakakibara2019,Nomura2019,Lechermann2020a,Lechermann2020b} have shown that the 3$d_{z^2}$ orbital is lower than the 3$d_{x^2-y^2}$ orbital and they predicted the presence of additional Fermi pockets at the $\Gamma$ and A points. There is a single Nickel-centered band of 3$d_{x^2-y^2}$ character crossing the Fermi level, and this band has a significant hole concentration due to self-doping from Nd-5$d$ states. 
This indicates an important role played by the low-lying Nd 5$d_{z^2}$ and 5$d_{xy}$ bands.

While the DFT bandstructure is well established, there is a general consensus about the strongly correlated nature of the material, which requires a full many-body description of a physically motivated low energy model. 
On the other hand, there is an intense debate about the number of Ni bands that need to be included in a realistic low-energy model. Some authors have argued that the nickelates are an almost perfect realization of a single-orbital Hubbard model,\cite{Nomura2019,Kitatani2020} while other groups have emphasized the modest splittings between the Ni 3$d$ bands, relative to the Hund coupling, the importance of high-spin configurations, and other multi-orbital aspects.\cite{Lee2004,Jiang2019,Werner2020,Lechermann2020b}

Also on the experimental side the new nickelate superconductor defies naive expectations: contrary to the cuprates, 
the undoped parent compound is weakly metallic and no evidence of an antiferromangetically ordered state has been found so far.\cite{Li2019,Li2020} This challenges the widespread assumption of a key role played by antiferromagnetic spin fluctuations in the unconventional pairing mechanism. The superconducting dome is similar to that of hole doped La$_2$CuO$_4$ but with a narrower doping window, a smaller T$_c\sim$10 K, and a double-peak structure.\cite{Li2020,Zheng2020} Both on the under- and overdoped side, superconductivity emerges from a bad metallic (weakly insulating) state. A recent spectroscopic study\cite{Goodge2020} supports the multi-orbital Mott-Hubbard picture, but also finds indications for the formation of Zhang-Rice singlets.  

\begin{figure}[t]
\includegraphics[width=1\textwidth]{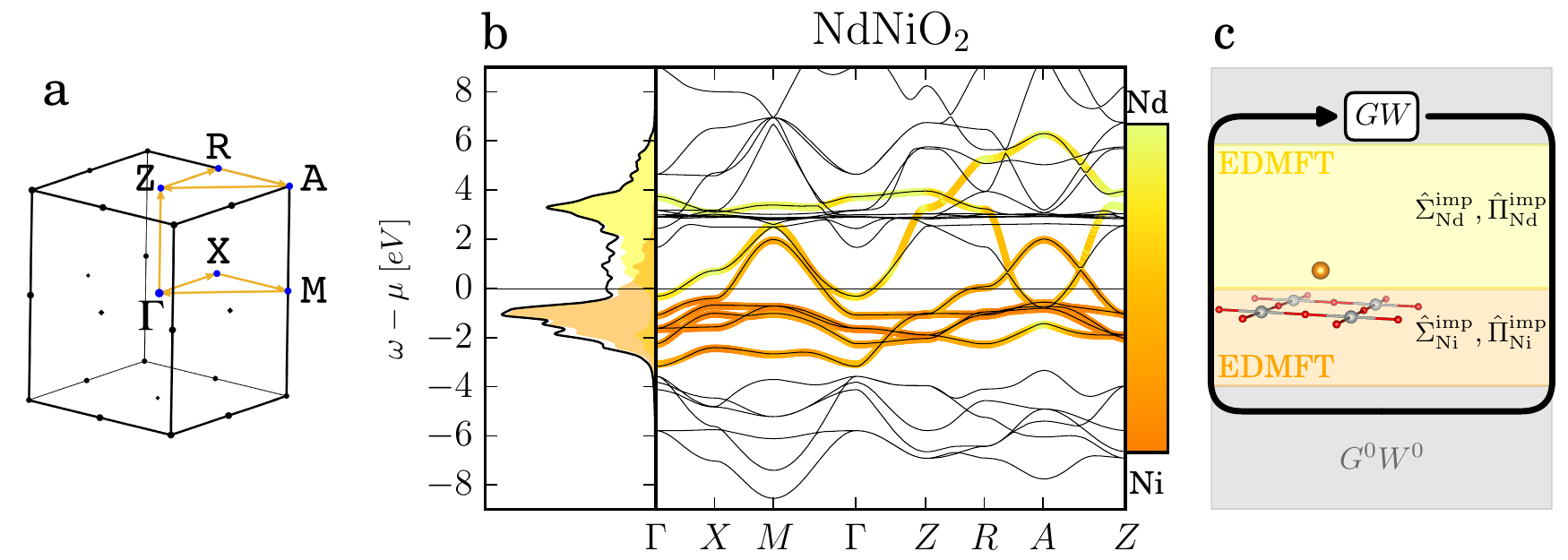}
\caption{{\bf Low-energy model and computational scheme.} {\bf a}, First Brillouin zone of NdNiO$_2$ and high-symmetry lines. {\bf b}, LDA bandstructure and low-energy space spanned by seven bands of predominantly Nd and Ni character. {\bf c}, Schematic representation of the multi-tier $GW$+EDMFT framework with two separate EDMFT calculations of local correlation and screening effects in the two Nd and five Ni orbitals, and nonlocal $GW$ contributions to the self-energies $\Sigma$ and polarizations $\Pi$. The initial downfolding to the low-energy space is performed with single-shot $G^0W^0$.  \label{Akw_scheme}}
\label{fig_bandstructure}
\end{figure}

Here we employ the recently developed multi-tier $GW$+EDMFT method,\cite{Boehnke2016,Nilsson2017,Petocchi2019} which enables an ab-initio simulation of strongly correlated materials {\it without adjustable parameters}, to clarify the electronic structure and the importance of multi-orbital physics in undoped and hole-doped NdNiO$_2$. For this purpose, we extend the $GW$+EDMFT method to two coupled interacting low energy models for Nickel and Neodymium containing five and two bands, respectively. This low-energy theory with self-consistently computed dynamically screened interaction parameters is embedded without double counting of interaction energies into an ab-initio bandstructure, as illustrated in Fig.~\ref{fig_bandstructure}. Our computational scheme starts with a DFT calculation in the local density approximation (LDA), where we employ the Virtual Crystal Approximation (VCA) to realistically account for the shifts of the bands induced by hole doping. The downfolding to the low-energy subspace hosting the Nd and Ni bands is achieved with a single-shot $G^0W^0$ calculation. In  this way, the high-energy degrees of freedom are incorporated via a frequency and momentum dependent self-energy and polarization into the bare propagators and bare interactions of the low-energy model. The latter is solved using self-consistent $GW$+EDMFT, with separate EDMFT impurity models for Nd and Ni.  The $GW$+EDMFT formalism\cite{Biermann2003} captures different aspects of interacting electron systems: $GW$ describes collective long-range charge fluctuations and dynamical screening effects, while EDMFT, the extension of DMFT to systems with nonlocal interactions,\cite{Sun2002} captures the effect of local Coulomb repulsions. These two techniques are combined by replacing the local part of the $GW$ self energy and polarization with the corresponding EDMFT estimates (see Methods). Both the Fermionic and Bosonic dynamical mean fields, represented respectively by the hybridization function and interaction tensor, are updated in the self-consistency loop,  resulting in a parameter-free ab-initio simulation, which only requires the definition of a physically motivated low-energy space. 

\begin{figure}
\includegraphics[width=0.8\textwidth]{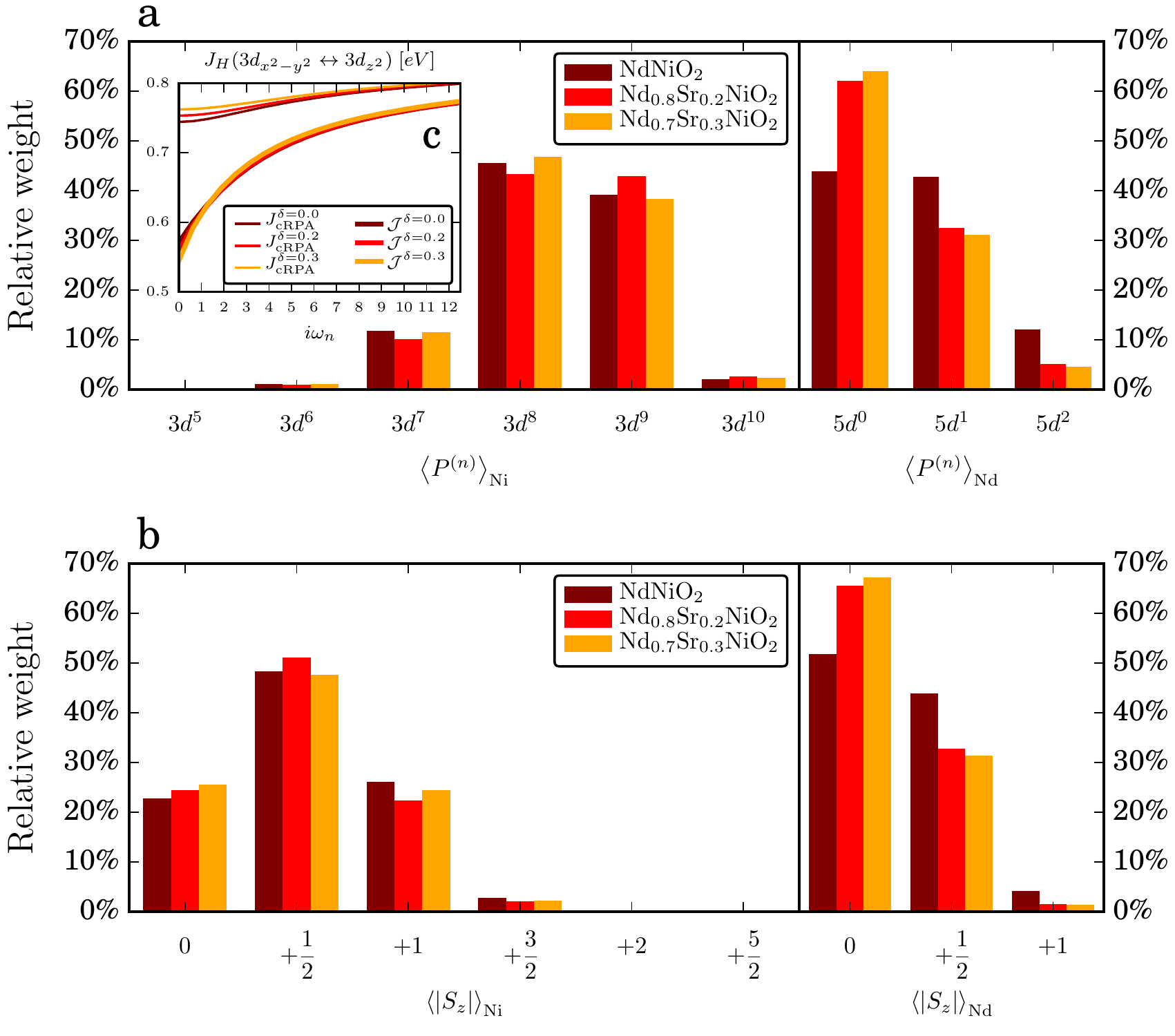}
\caption{{\bf Occupation and spin statistics for Ni and Nd.} {\bf a}, Probability distributions for the Ni (left) and Nd (right) atoms to be in the indicated charge states. {\bf b}, Probability distributions for the absolute value of the $z$-component of the spin. The different colors correspond to different hole dopings (dark red: undoped, red: close to optimal doping, orange: overdoped). {\bf c}, Frequency and doping dependence for the Hund coupling between Ni-3$d_{x^2-y^2}$ and Ni-3$d_{z^2}$. \label{statistics}}
\end{figure}

A useful feature of this EDMFT based method is that it gives access to the relevant occupation and spin states on Ni and Nd. In Fig.~\ref{statistics}{\bf a} we present the corresponding histograms for different hole dopings. Even in the undoped compound, where the total density is fixed at the LDA nominal value, the configuration probabilities $P^{(n)}$, which measure the fraction of time spent in the indicated atomic configurations, clearly reveal the multi-orbital behavior of the Nickel subsystem: the 3$d^8$ states are relevant and comparable in magnitude to the naively expected 3$d^9$ states, and also the charge fluctuations to the 3$d^7$ states cannot be ignored. This is largely due to the self-doping from Nd, which accommodates up to two electrons. Interestingly, the configuration probabilities of Ni vary with the dopant concentration in a nonmonotonic fashion. In agreement with the results from DFT+$U$ studies,\cite{Choi2020} we find that the doped holes essentially end up on the Nd sites, as can be concluded from the monotonous increase in the 5$d^0$ configuration. The nonmonotonic behavior observed on the Ni sites, which have an electron density pinned at 8.3 (see also the Supplementary Material (SM)), results from doping-induced changes in the hybridization strength and the self-consistently computed interaction parameters. In the following matrices we report, for the undoped and the optimally doped compounds, the screened values of the effective Ni onsite interaction $\mathcal{U}(\omega_n=0)$ in the upper triangular section and the effective Hund coupling $\mathcal{J}(\omega_n=0)$ in the lower one: 
\begin{align}
\hat{\mathcal{I}}_{\mathrm{Ni}}^{\delta=0.0}=\begin{array}{c}
\begin{array}{c|ccccc}
 & d_{xz} & d_{yz} & d_{xy} & d_{z^{2}} & d_{x^{2}-y^{2}}\\
\hline d_{xz} & 4.95 & 3.50 & 3.39 & 3.82 & 3.08\\
\cline{2-2}d_{yz} & \multicolumn{1}{c|}{0.69} & 4.95 & 3.39 & 3.82 & 3.08\\
\cline{3-3}d_{xy} & 0.69 & \multicolumn{1}{c|}{0.69} & 4.74 & 3.07 & 3.54\\
\cline{4-4}d_{z^{2}} & 0.50 & 0.50 & \multicolumn{1}{c|}{0.71} & 4.85 & 2.82\\
\cline{5-5}d_{x^{2}-y^{2}} & 0.53 & 0.53 & 0.35 & \multicolumn{1}{c|}{0.57} & 3.98
\end{array}\end{array},\;\;
\hat{\mathcal{I}}_{\mathrm{Ni}}^{\delta=0.2}=\begin{array}{c|ccccc}
 & d_{xz} & d_{yz} & d_{xy} & d_{z^{2}} & d_{x^{2}-y^{2}}\\
\hline d_{xz} & 5.24 & 3.81 & 3.62 & 4.18 & 3.29\\
\cline{2-2}d_{yz} & \multicolumn{1}{c|}{0.68} & 5.24 & 3.62 & 4.18 & 3.29\\
\cline{3-3}d_{xy} & 0.67 & \multicolumn{1}{c|}{0.67} & 4.85 & 3.35 & 3.66\\
\cline{4-4}d_{z^{2}} & 0.49 & 0.49 & \multicolumn{1}{c|}{0.70} & 5.32 & 3.09\\
\cline{5-5}d_{x^{2}-y^{2}} & 0.51 & 0.51 & 0.34 & \multicolumn{1}{c|}{0.56} & 4.07
\end{array}.
\label{eqn:Uparams}
\end{align}
In particular, we notice that the $\mathcal{J}$ values decrease with hole doping (see also Fig.~\ref{statistics}{\bf c}), while the inter-orbital interactions increase more strongly than the intra-orbital ones. 

In a strongly correlated, half-filled single-band model one would expect to find spins with magnitude $|S_z|$ close to one half, and fluctuations to states with $|S_z|=0$. As can be seen from Fig.~\ref{statistics}{\bf b}, in our seven-orbital model, fluctuations to both high and low spin states are significant. The high-spin states are stabilized by the Hund coupling. Even if the doping-dependent changes on the Ni ion are small, the non monotonicity in the charge and spin statistics indicates that the optimally doped compound is closest to the behavior one would expect from a model containing a single $d_{x^2-y^2}$ band, and that the expected increase in the population of spin-1 moments with hole doping\cite{Lee2004,Jiang2019,Werner2020} only sets in on the overdoped side of the experimental $T_c$ dome. 

\begin{figure}
\includegraphics[width=1\textwidth]{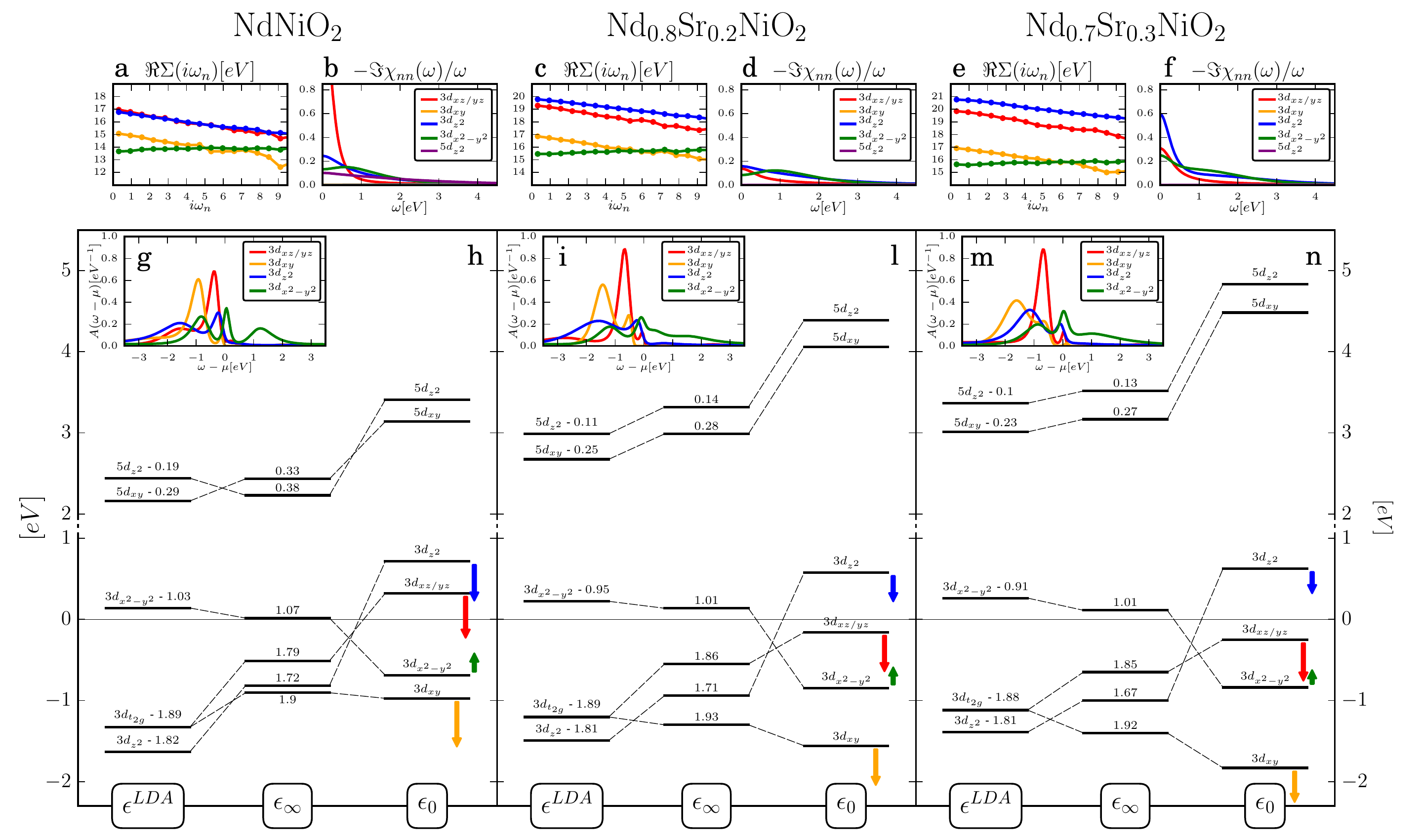}
\caption{{\bf Level diagrams and energy dependent shifts of the accessible states.} The results for indicated dopings are summarized in the main panels {\bf h}, {\bf l}, {\bf n}. The level diagrams on the left, marked with $\epsilon^{LDA}$, show the average energies of the DFT bands. The level diagrams in the middle, marked $\epsilon_\infty$, indicate the center of mass of the correlated DOS, which is plotted in the insets (panels {\bf g}, {\bf i}, {\bf m}). 
The numbers near the levels report the orbital occupations. 
The right level diagrams, marked $\epsilon_0$, indicate the level energies of the EDMFT impurity problems, shifted by $\Re\Sigma\left(\omega_n=0\right)$. To illustrate the effect of the $\omega_n$-dependence of the self-energy (shown in panels {\bf a}, {\bf c}, {\bf e}), we illustrate by the arrows the level shift produced by increasing $\omega_n$ to $3$ eV (except for the green arrow, where we used $10$ eV for a better visualization of the shift direction). Panels {\bf b}, {\bf d}, {\bf f} show the charge susceptibilites for different orbitals on the real-frequency axis. \label{levels}}
\end{figure}
 
The multi-orbital nature of the undoped and hole-doped nickelate compounds is the consequence of local energy renormalizations induced by the local EDMFT self-energies. In particular we find a strong frequency dependence of the real part of $\hat{\Sigma}^{\mathrm{Ni}}_{\mathrm{loc}}$, which is positive and (except for the $d_{x^2-y^2}$ orbital) increasing with decreasing energy. As a consequence, local energies which correspond to fully occupied orbitals at high energy, become available for low energy hole-like charge fluctuations (or fast virtual charge excitations), in agreement with the previously discussed configurational statistics. To demonstrate this effect of the self-energy we present in Fig.~\ref{levels} the local energies obtained using three different approaches. We first consider the diagonal entries of the real space LDA Hamiltonian at the origin, $\epsilon^\text{LDA}=\mathcal{H}^\text{LDA}\left(\mathbf{R}=0\right)-\mu$. In agreement with the DFT results reported in the literature,\cite{Sakakibara2019,Nomura2019} this yields a picture compatible with a single-band description. To represent the local energies of the interacting system, we consider the center of mass of the local spectral function, $\epsilon_{\infty}=\int\omega A(\omega)d\omega$. (This includes all the self-energy terms connecting the different tiers, while just adding the real part of the local EDMFT self-energy to $\epsilon^\text{LDA}$ would for example miss the $GW$ contributions.) Also in this picture, which may be thought of as the Hartree limit of our result, the Ni states are mostly occupied, except for the 3$d_{x^2-y^2}$ orbital. Finally, to illustrate the effect of the frequency dependence of $\Re\hat{\Sigma}^{\mathrm{Ni}}_{\mathrm{loc}}(\omega_n)$, we recall that the fermionic Weiss field $\hat{\mathcal{G}}^{-1}=i\omega_{n}+\mu-E_\text{loc}-\hat{\Delta}\left(\omega_{n}\right)$ (with $\hat{\Delta}\left(\omega_{n}\right)$ the hybridization function) is defined with respect to an effective local energy $E_\text{loc}$ determined by the self-consistency equations. In particular, $E_\text{loc}$ incorporates the modifications of the bandstructure in the downfolding and the $\mathbf{k}$-dependent $GW$ contributions to the self-energy. In the EDMFT impurity calculation, the local self-energy is then added by the solver to this effective level. We thus plot in the right hand level diagrams the renormalized impurity level positions $\epsilon_{0}=E_\text{loc}-\mu+\Re\Sigma(0)$. The result indicates a substantial shift of the 3$d_{z^2}$ and 3$d_{xz,yz}$ orbitals to higher energies, compared to the Hartree limit, while the 3$d_{x^2-y^2}$ and 3$d_{xy}$ orbitals shift to lower energies. The arrows in the level diagrams demonstrate that increasing the frequency $\omega_n$, where the self-energy is evaluated, to $3$ eV (approximate bandwidth) brings the levels closer to their $\epsilon_{\infty}$ values. In recent LDA+DMFT studies, Lechermann\cite{Lechermann2020a,Lechermann2020b} emphasized the behaviour of the Ni-3$d_{z^2}$ orbital, which for his large value of the onsite interaction ($U=10$ eV) empties out and thus enables the system to undergo an orbital selective Mott transition. Even though our selfconsistently computed static onsite interactions are a factor of two smaller, the result in Fig.~\ref{levels} is reminiscent of this phenomenology. 
The same tendency can be seen in the Ni-3$d_{z^2}$ dispersion of the interacting system (see SM Figs.~6-8) in which the flat part of the  band is shifted up in energy very close to the Fermi level. 
As a further support of our interpretation, we plot in Fig.~\ref{levels}({\bf b}, {\bf d}, {\bf f}) the Fourier transforms of the local charge susceptibilities $\hat \chi_{nn}(\tau)=\left\langle \hat{n}(\tau)\hat{n}(0)\right\rangle$ on the real frequency axis. This quantity defines the screening due to local charge fluctuations and the contribution of each orbital is easily identifiable. As expected no contribution is found from Ni-3$d_{xy}$ and Nd-5$d_{xy}$ which are, respectively, completely filled and completely empty, while Nd-5$d_{z^2}$ contributes only in undoped NdNiO$_2$ which hosts the hole pocket at the $\Gamma$ point. Ni-3$d_{xz/yz}$ fluctuates strongly in the undoped compound, in agreement with the energy-dependent shift of the effective level position (red arrow). Overall, the optimally doped compound is least affected by charge fluctuations, which suggests that the latter do not play a role in the pairing mechanism. 

\begin{figure}
\includegraphics[width=0.6\textwidth]{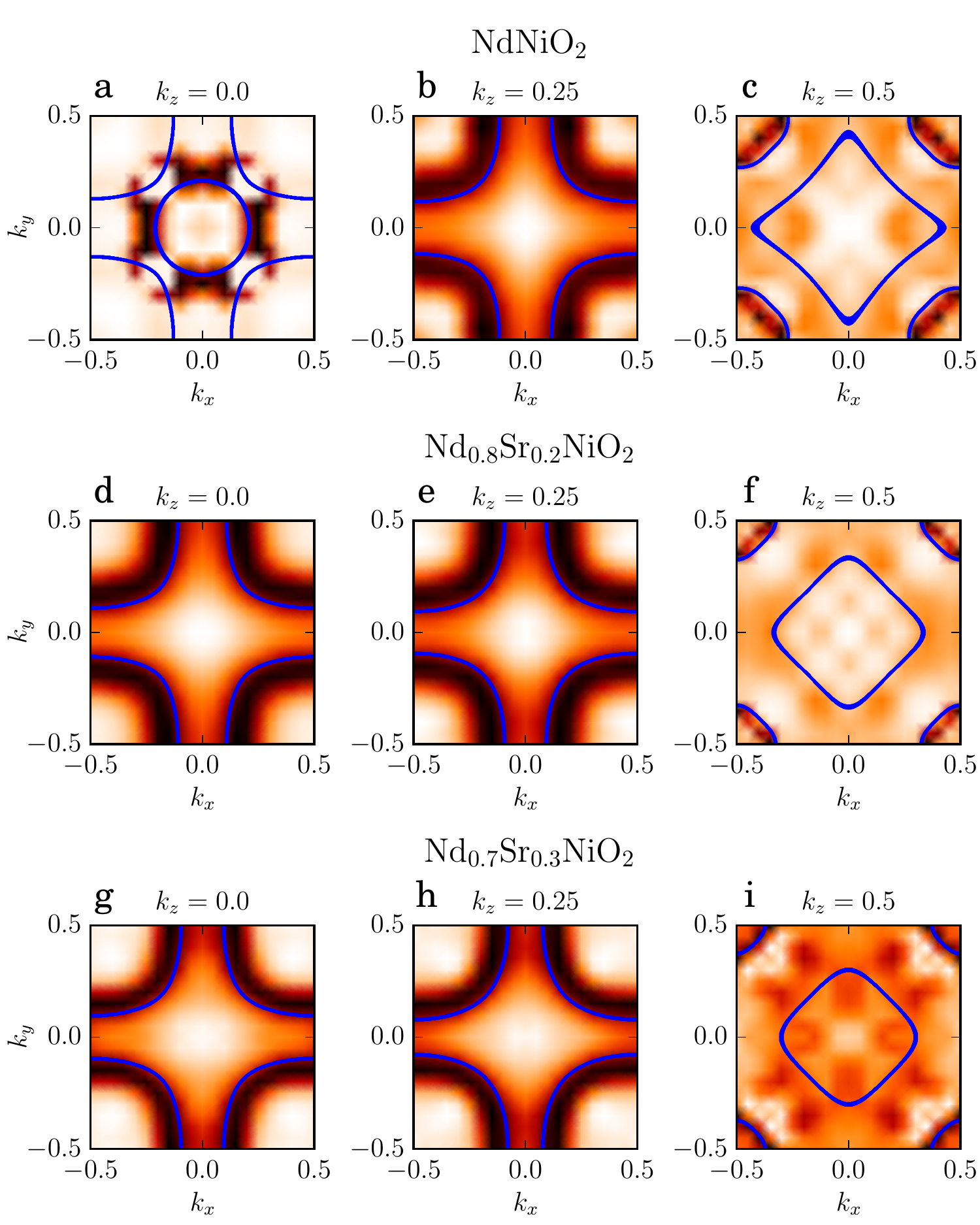}
\caption{{\bf Doping evolution of the Fermi surface.} The blue lines show the DFT Fermi surfaces, while the intensity plots for the occupied states illustrate the ``Fermi surfaces" of the interacting systems, calculated as the total spectral function at $\omega=0$. Here, the intensity scale is relative to the maximum value of this function. Analogous orbital-resolved plots with a fixed intensity scale can be found in the SM. \label{fermi}}
\end{figure}

In Fig.~\ref{fermi} we show how the tight-binding Fermi surfaces are modified by the interactions. The interacting result corresponds to the trace of the $\mathbf{k}$-resolved spectral function evaluated at $\omega=0$. In agreement with the existing literature,\cite{Sakakibara2019,Nomura2019,Lechermann2020a,Choi2020} our non-interacting reference system contains two hole pockets centered at the A and $\Gamma$ points, associated with Nd-5$d_{xy}$ and Nd-5$d_{z^2}$ states, respectively. DFT predicts that this latter band is continuously pushed above the Fermi level as the hole doping is increased. On top of this shift our results, already at the  $G^0W^0$ level, indicate a substantial flattening of the bands along the $\Gamma$-X and $\Gamma$-M directions (see SM and sketch of the Brillouin zone in Fig.~\ref{Akw_scheme}{\bf a}). Due to the very low carrier concentration there are little additional deformations induced by the local interactions. The difference between the $G^0W^0$ and $GW$+EDMFT treatment concerns mainly an increase in the local energy. The combination of these two effects yields a broadened $\Gamma$ pocket, which has the highest intensity at $k_z=0$ in the undoped compound (see Fig.~\ref{fermi}{\bf a}). Moving to the intermediate value of $k_z=0.25$ (see Fig.~\ref{fermi}b), in addition to the dominant Ni-3$d_{x^2-y^2}$ contribution, we notice some spectral weight in the M-A direction (at the corners of the Brillouin zone), which has a Ni-3$d_{xz,yz}$ character. This feature is consistent with the charge susceptibility results for undoped NdNiO$_2$. Finally, at  $k_z=0.5$, we find significant deviations from the DFT result. The Ni-3$d_{x^2-y^2}$ pocket centered at the Z point gives way to a Ni-3$d_{z^2}$ feature, as can be seen from the lobes emerging in the Z-R direction. In this region, the dispersion is already flat at the DFT level but, as several of our results suggest, the interaction effects significantly increase the local energy. Upon doping, the $\Gamma$ pocket and the small M-A weight are removed, leaving the typical $d_{x^2-y^2}$ shape at $k_z=0,0.25$, while the lobes originating from Ni-3$d_{z^2}$ states persist. The latter  indicates an active role of these states at the Fermi level in all the studied setups. A general observation is that the Fermi surface, and especially the Ni-3$d_{x^2-y^2}$ contribution, gets closer to the LDA result with increasing hole doping, which is consistent with the system becoming more metallic and less correlated.

\begin{figure}
\includegraphics[width=1\textwidth]{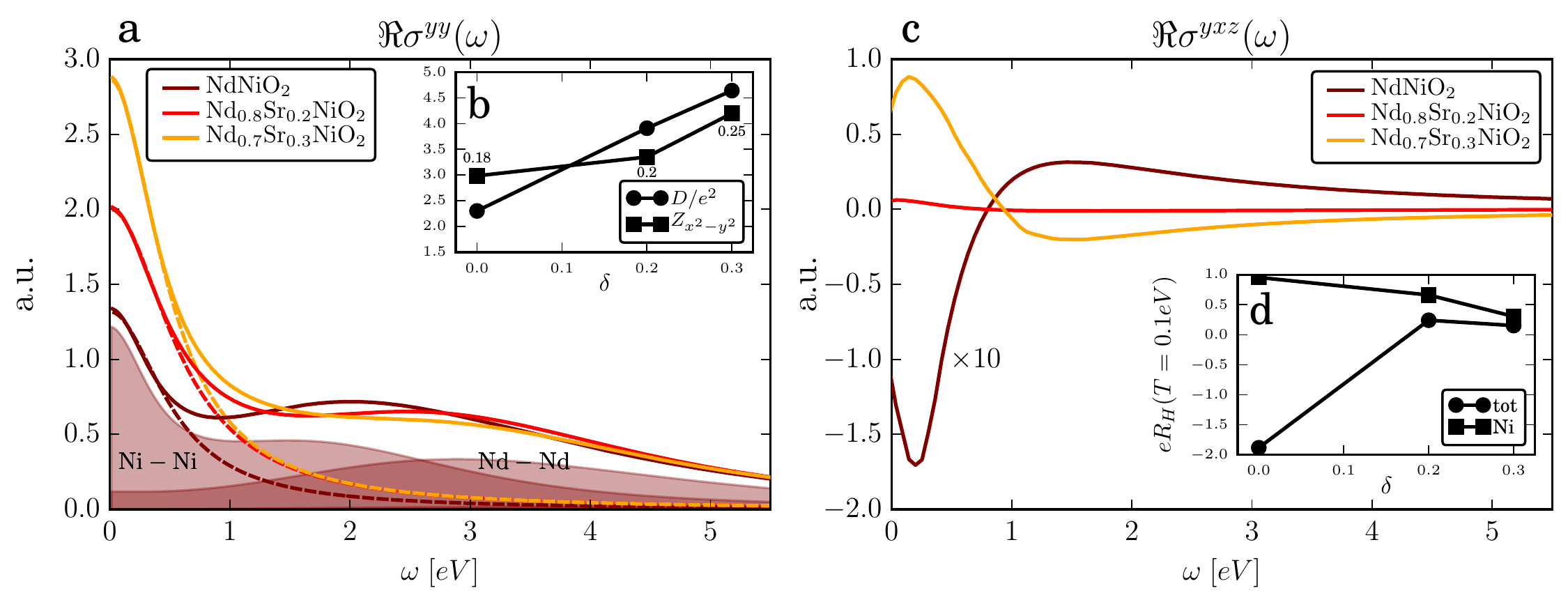}
\caption{{\bf Linear response conductivities.} {\bf a}, The optical conductivity as a function of doping. The shading indicates the contribution of the two sites in the undoped compound. {\bf b}, Drude weight as a function of doping and comparison to the DMFT estimate of the mass enhancement for the Ni-3$d_{x^2-y^2}$ orbital (derivative of the local self-energy), with values indicated near the points. {\bf c}, Hall conductivity as a function of doping. The curve of the undoped system has been rescaled by a factor of 10 for better visibility. {\bf d}, Hall coefficient for the whole system and restricted to the Ni site only. \label{optconds} }
\end{figure}

In Fig.~\ref{optconds} we present results for the optical and Hall conductivities computed for all the doping levels within linear response theory. In the limit of infinite dimensions, the irreducible vertex function becomes local and it can be neglected when the current vertex is odd with respect to the $\mathbf{k}$ vector.\cite{Pruschke1993,note1}  As a consequence, the Kubo formula for the conductivities requires the evaluation of bubble-like diagrams involving only the Green's functions and current vertices. The response function parallel to an electric field in the $y$ direction is the optical conductivity $\Re\sigma^{yy}(\omega)$. In interacting systems it features two contributions, a Drude-like peak at low frequency, originating from excitations within the quasiparticle band, and incoherent structures $\sigma^\text{inc}(\omega)$ at higher energies stemming either from inter-band excitations or, for a correlated band, from excitations to the Hubbard satellites:\cite{Jarrell1995}
\begin{align}
\Re\sigma^{yy}\left(\omega\right)=\frac{D}{\pi}\frac{\tau}{1-\left(\omega\tau\right)^{2}}+\sigma^\text{inc}\left(\omega\right),
\label{eq_drude}
 \end{align}
where $\tau$ denotes the relaxation time and $D=e^{2}\frac{n}{m^{*}}$ is the Drude weight written in terms of the carrier density $n$ and effective mass $m^*$. Our results in Fig.~\ref{optconds}{\bf a} capture both the low and high energy peaks with weight that is transferred from the latter to the former as a function of hole doping. By separately computing the contributions from the Nd and Ni sites, we find that Nd yields a high energy peak which is essentially fixed at $\omega\sim3.5$ eV, regardless of the doping concentration, so that the spectral weight transfer originates from Ni (see SM for the site-resolved $\Re\sigma^{yy}$ at various doping levels). These observations demonstrate a more metallic behaviour of the Ni sub-system with increasing hole-doping, despite the almost constant occupation and the increase in the interaction parameters (Eq.~\eqref{eqn:Uparams}). The weaker correlations result from the stronger increase in the {\it interorbital} interactions, compared to the intraorbital interactions, and the corresponding weakening of the Hund couplings, and hence are a nontrivial manifestation of multi-orbital effects. To quantify the degree of metallicity of the correlated multiorbital system, we extract the Drude weight by fitting the low energy part of $\Re\sigma^{yy}(\omega)$ to Eq.~\eqref{eq_drude}. This weight is inversely proportional to an effective mass defined for the entire system and is shown in Fig.~\ref{optconds}{\bf b}. The result indicates that doping indeed brings the system into a more metallic, less correlated state. This is consistent with several measurements \cite{Li2020,Zheng2020,Goodge2020} which report a decrease in the resistivity with hole doping on the underdoped side of the $T_c$ dome, which appears to be least affected by disorder. In Fig.~\ref{optconds}{\bf b} we also plot the quasiparticle weight obtained from the local self-energy of the Ni-3$d_{x^2-y^2}$ orbital. Both estimates are in qualitative agreement with the DFT+DMFT results reported by Kitatani {\it et al.}\cite{Kitatani2020}.

While for the standard conductivity one needs to take the derivative of the free energy with respect to the vector potential twice, the Hall conductivity $\Re\sigma^{yxz}$ involves a third order process, which requires an additional vertex insertion. In deriving the Hall current-current correlator we extended (specifically for the  $GW$+EDMFT implementation) the approach described in Ref.~\onlinecite{Voruganti1992} to the multiorbital case, as explained in the SM.  The Hall coefficient, defined as
\begin{align}
R_{H}=\frac{\Re\sigma^{yxz}(0)}{\Re\sigma^{yy}(0)\Re\sigma^{xx}(0)},
\end{align}
characterizes the nature of the charge carriers: it has a negative sign for electron-like Fermi surfaces while it is positive for hole-like ones. Our results for $R_{H}$, in units of the inverse electron density, are reported in Fig.~\ref{optconds}{\bf d} and are in qualitative agreement with the recent experimental findings of Li {\it et al.}\cite{Li2020} and Zheng {\it et al.}\cite{Zheng2020}.

\noindent
{\bf Discussion}

To address the physics of doped NdNiO$_2$ in the normal state, and in particular the self-doping effect, we extended the recently developed $GW$+EDMFT approach to multi-site systems. This method has the significant advantage of being free from ill-defined double countings or arbitrary choices of interaction parameters. It captures the dynamical screening due to long ranged and local charge fluctuations and self-consistently computes the local and nonlocal interactions appropriate for the low-energy model. Our results allow us to assert with confidence that undoped and hole doped NdNiO$_2$ represent genuine multi-orbital systems. Orbitals that, within a DFT description, are expected to be fully occupied and essentially inert, are lifted closer to the Fermi level by the interactions. As a consequence, also other orbitals than the naively expected Ni-3$d_{x^2-y^2}$ become involved in low-energy charge fluctuations, and in the formation of high-spin states. We find prominent contributions from $d^8$ and $|S_z|=1$ local configurations, even in the undoped compound. The evolution of the metallicity with hole doping is a direct consequence of the doping-induced changes in the dynamically screened inter-orbital interactions. Therefore, a low-energy model consisting of a single $d$ orbital with fixed and static interactions misses important aspects of the physics of undoped and hole-doped NdNiO$_2$. On the other hand the high frequency (Hartree) limit of our approach looks consistent with the effective single-band  picture, and in this sense our results reconcile the two fronts of the debate. As a nontrivial check of our parameter-free ab-initio formalism, we computed the optical and Hall conductivities within linear response theory and found qualitative agreement both with the reported decrease in the resistivity in the doping range least affected by disorder,\cite{Li2020} and with the recently measured Hall coefficients.\cite{Li2020,Zheng2020} 
\\
\\
\noindent
{\bf Methods}
\label{sec_method_end}

Our simulations are based on the parameter-free multi-tier $GW$+DMFT scheme,\cite{Boehnke2016,Nilsson2017,Petocchi2019} which treats different energy scales with appropriate levels of accuracy, 
and which self-consistently  computes dynamically screened interactions for the low-energy model space. 

The initial DFT calculations are performed using the full-potential linearized augmented plane-wave code \textit{FLEUR},\cite{Fleurcode} with the experimental lattice parameters for NdNiO$_2$ ($a=b=3.92$ \AA, $c=3.28$ \AA \cite{Hayward2003}), and a $16 \times 16 \times 16$ $k$-point grid. To treat the $4f$-states, which appear unphysically close to the Fermi energy,  we use a manual core set-up with the $4f^3$ electrons in the core. To correct also the bands originating from the unoccupied $4f$ states we use a self-consistent LDA+$U$+cRPA (constrained random phase approximation) approach, where the cRPA $U$ \cite{Aryasetiawan2004} calculated for the unoccupied $4f$ bands is used in a new LDA+$U$ calculation iteratively until convergence is reached. To simulate the doped compound we use the VCA, where the doping is achieved by replacing a fraction of Nd with Pr, instead of the experimentally used Sr. This is due to a technical limitation, which only allows to dope with consecutive elements. The low-energy model is defined using maximally-localised Wannier functions obtained from the Wannier90 library, \cite{Marzari1997,Mostofi2008,Freimuth2008,Sakuma2013} while the $G^0W^0$ calculations are performed using the \textit{SPEX} code \cite{Friedrich2010} with $8\times 8\times 8$ $k$-points and 250 bands.

Within the low-energy space with two Nd and five Ni orbitals, we perform a self-consistent $GW$+EDMFT calculation. In this approach local self-energies $\hat{\Sigma}^{\mathrm{imp}}$ and polarizations $\hat{\Pi}^{\mathrm{imp}}$ for Nd and Ni are computed from separate EDMFT impurity problems with self-consistently optimized fermionic and bosonic Weiss fields $\mathcal{G}$ and $\mathcal{U}$, respectively. To these local impurity self-energies and polarizations, we add the nonlocal $GW$ components, 
\begin{align}
	\Sigma_{ab}^{\mathrm{nonloc}}(\mathbf{q},\tau)=	-\sum_{\mathbf{k}cd}G_{cd}(\mathbf{k},\tau)W_{acbd}(\mathbf{q}-\mathbf{k},\tau)+\sum_{cd}G_{cd}^{\mathrm{loc}}(\tau)W_{acbd}^{\mathrm{loc}}(\tau) , \\
	\Pi_{acbd}^{\mathrm{nonloc}}(\mathbf{q},\tau)=	\sum_{\mathbf{k}}G_{ab}(\mathbf{k},\tau)G_{dc}(\mathbf{k}-\mathbf{q},-\tau)-G_{ab}^{\mathrm{loc}}(\tau)G_{dc}^{\mathrm{loc}}(-\tau),
\end{align}
where $W_{acbd}$ denotes the elements of the screened interaction, and $G_{ab}$ those of the interacting Green's function. The $GW$+EDMFT cycle starts from an initial guess for $\hat{\Sigma}^{\mathrm{imp}}$  and $\hat{\Pi}^{\mathrm{imp}}$.  Then, given the noninteracting lattice Hamiltonian $\hat{\mathcal{H}}(\mathbf{k})$ for the localized Wannier orbitals of the low-energy space, non-local $GW$ self-energies and polarizations are computed. The sum of these two contributions yields the momentum-dependent self-energy $\Sigma_{\mathbf k}$ and polarization $\Pi_{\mathbf q}$ entering the lattice sums,
\begin{align}
	\hat{G}^\mathrm{loc}=	\frac{1}{N}\sum_\mathbf{k} \big((\hat{G}^{(0)}_\mathbf{k})^{-1}-\hat{\Sigma}_\mathbf{k}\big)^{-1}, \label{eq:Gloc} \\
	\hat{W}^\mathrm{loc}=\frac{1}{N} \sum_\mathbf{q}\hat{U}_\mathbf{q}\left(\mathbbm{1}-\hat{\Pi}_\mathbf{q} \hat{U}_\mathbf{q}\right)^{-1}, \label{eq:Wloc} 
\end{align}
where $\hat{G}^{(0)}_\mathbf{k}$ is the noninteracting propagator of the low energy subspace, which incorporates also the $G^0W^0$ contribution (see Eq.~\eqref{eqn:fullG} below), and $\hat{U}_\mathbf{q}$ the ``bare" Coulomb interaction resulting from the initial $G^0W^0$ downfolding. These local Green's functions and screened interactions are obtained by inversion in the full (7$\times$7) orbital space. The EDMFT self-consistency condition then demands that the projections of $\hat{G}^\mathrm{loc}$ and $\hat{W}^\mathrm{loc}$ onto the Ni and Nd sites are equal to the impurity Green's functions and screened interactions for these sites. Using these $G^\text{imp}$ and $W^\text{imp}$ and the impurity self-energies and polarizations, the fermionic and bosonic Weiss fields are computed as
\begin{align}
	\hat{\mathcal{G}}_i=\left(\hat{\Sigma}_i^\mathrm{imp}+(\hat{G}_i^\mathrm{imp})^{-1}\right)^{-1} & \qquad i\in\left\{ \mathrm{Ni},\mathrm{Nd}\right\}  , \label{eq:GWeiss}\\
	\hat{\mathcal{U}}_i=\hat{W}_i^\mathrm{imp}\left(\mathbbm{1}+\hat{\Pi}_i^\mathrm{imp}\hat{W}_i^\mathrm{imp}\right)^{-1} & \qquad i\in\left\{ \mathrm{Ni},\mathrm{Nd}\right\},    \label{eq:WWeiss} 
\end{align}
and used as inputs for the two EDMFT impurity problems. 
(In these inversions, all orbital indices are restricted to either the Ni or Nd sites.)
An efficient continuous-time Monte Carlo solver\cite{Werner2006,Hafermann2013} for models with dynamically screened interactions\cite{Werner2010} then yields the impurity Green's functions $\hat G^\text{imp}_i$ and density-density correlation functions $\hat\chi^\text{imp}_i$.  After the Fourier transformation of $\hat G^\text{imp}_i$ and  $\hat\chi^\text{imp}_i$, the new $\hat{\Sigma}^{\mathrm{imp}}_i$  and $\hat{\Pi}^{\mathrm{imp}}_i$ are computed from the Weiss fields, the impurity Green's functions and charge susceptibilities:
\begin{align}
	\hat{\Sigma}_i^\mathrm{imp}=\hat{\mathcal{G}}_i^{-1}-(\hat{G}_i^\mathrm{imp})^{-1} & \qquad i\in\left\{ \mathrm{Ni},\mathrm{Nd}\right\}  ,  \label{eq:Gsc}  \\
	\hat{\Pi}_i^\mathrm{imp}=\hat{\chi}_i^\mathrm{imp}\left(\hat{\mathcal{U}}_i\hat{\chi}_i^\mathrm{imp}-\mathbbm{1}\right)^{-1} & \qquad i\in\left\{ \mathrm{Ni},\mathrm{Nd}\right\}  \label{eq:Wsc}. 
\end{align}
Since all the self-energy and polarization contributions in this multi-tier scheme are diagrammatically defined, we can connect the different subspaces in a consistent way, without any double countings. The explicit expressions for the interacting lattice Green's functions and screened interactions are:
\begin{align}
G_{\mathbf{k}}^{-1}=&\overbrace{\mathrm{i}\omega _{n}+\mu -\varepsilon _{\mathbf{k}}^{\mathrm{DFT}}+V_{\mathrm{XC},\mathbf{k}}-\left(\Sigma_{\mathbf{k}}^{G^{\, 0}W^{\, 0}}-\Sigma_{\mathbf{k}}^{G^{\, 0}W^{\, 0}}\big|_{I}\right)}^{\text{LDA}+G^{0}W^{0},\; G_{I, \mathbf{k}}^{\,0}{}^{-1}}\notag\\
&\underbrace{-\left(\Sigma _{\mathbf{k}}^{GW}\big|_{I}-\Sigma^{GW}\big|_{C,\mathrm{loc}} + \Delta V_H|_{I} \right)}_{GW}\underbrace{-\Sigma ^{\mathrm{EDMFT}}\big|_{C,\mathrm{loc}}}_{\text{EDMFT}}\;,\label{eqn:fullG}\\
W_\mathbf{q}^{-1}=&\overbrace{v_\mathbf{q}^{-1}-\left(\Pi^{G^{\,0}G^{\, 0}}_\mathbf{q}-\Pi^{G^{\, 0}G^{\,0}}_\mathbf{q}\big|_I\right)}^{\text{cRPA}+G^{0}W^{0},\; U_{I,\mathbf{q}}^{-1}}\notag\\
&\underbrace{-\left(\Pi_\mathbf{q}^{GG}\big|_I-\Pi^{GG}\big|_{C,\mathrm{loc}}\right)}_{GW}\underbrace{-\Pi^\mathrm{EDMFT}\big|_{C,\mathrm{loc}}}_{\text{EDMFT}},\label{eqn:fullW}
\end{align}
where $C$ refers to the strongly correlated local subspaces (treated with EDMFT) and $I$ to the full 7-orbital subspace in which the $GW$ calculation is performed. $\Delta V_H$ is the Hartree contribution to the self-energy, $v_{\mathbf q}$ is the bare interaction and $V_\text{XC}$ the exchange-correlation potential (which is replaced by the $G^0W^0$ self-energy). A detailed derivation of the multi-tier GW+EDMFT approach can be found in Refs.~\onlinecite{Nilsson2017,Petocchi2019}.
%
%
%
%
\clearpage
\noindent
{\bf Acknowledgments}
FP, VC and PW acknowledge support from the Swiss National Science Foundation 
through NCCR MARVEL and the European Research Council through ERC Consolidator 
Grant 724103. FN and FA acknowledge financial support from the Swedish Research Council (VR).
The calculations were performed on the Beo04/Beo05 cluster at the University of Fribourg. We thank F. Lechermann, A. J. Millis, and H. Y. Hwang for helpful discussions. 
%
%
%
%

%
%
%
%

\begin{figure}[b]
\begin{centering}
\includegraphics[scale=0.9]{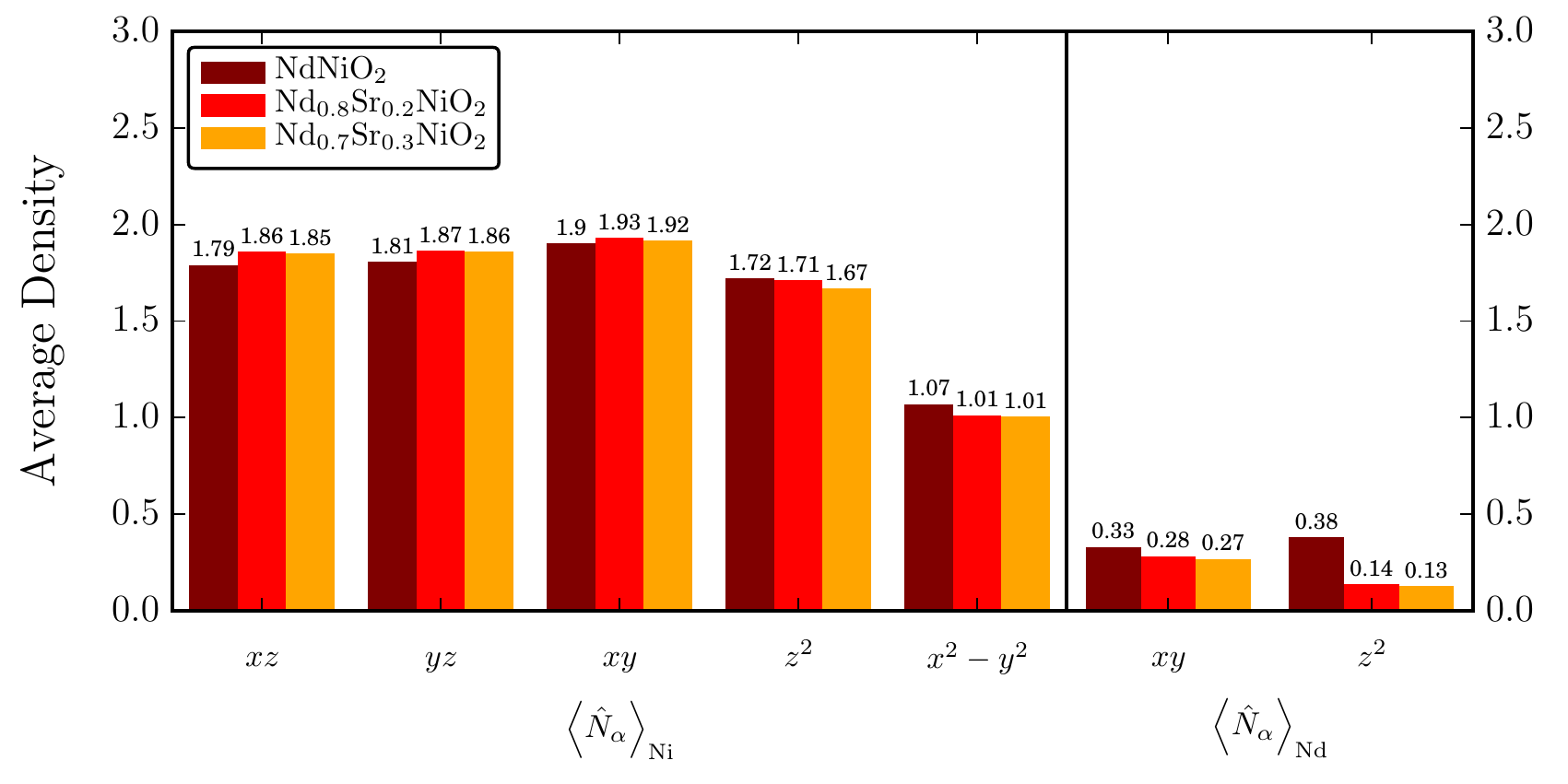}
\par\end{centering}
\caption{Density expectation values for Ni and Nd.}\label{density}
\end{figure}

\begin{figure}
\begin{centering}
\includegraphics[scale=0.9]{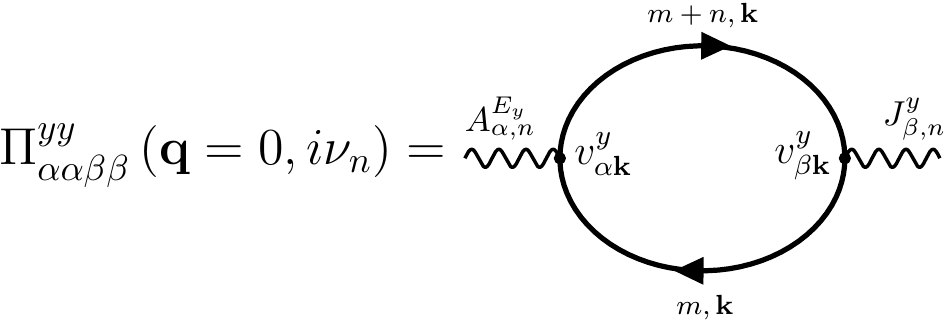}
\par\end{centering}
\caption{Second order diagrams contributing to the optical conductivity for the electric field and the induced current in the $y$ direction. 
Greek subscripts indicate the orbital index.}\label{diagram_optc}
\end{figure}

\begin{figure}
\begin{centering}
\includegraphics[scale=0.75]{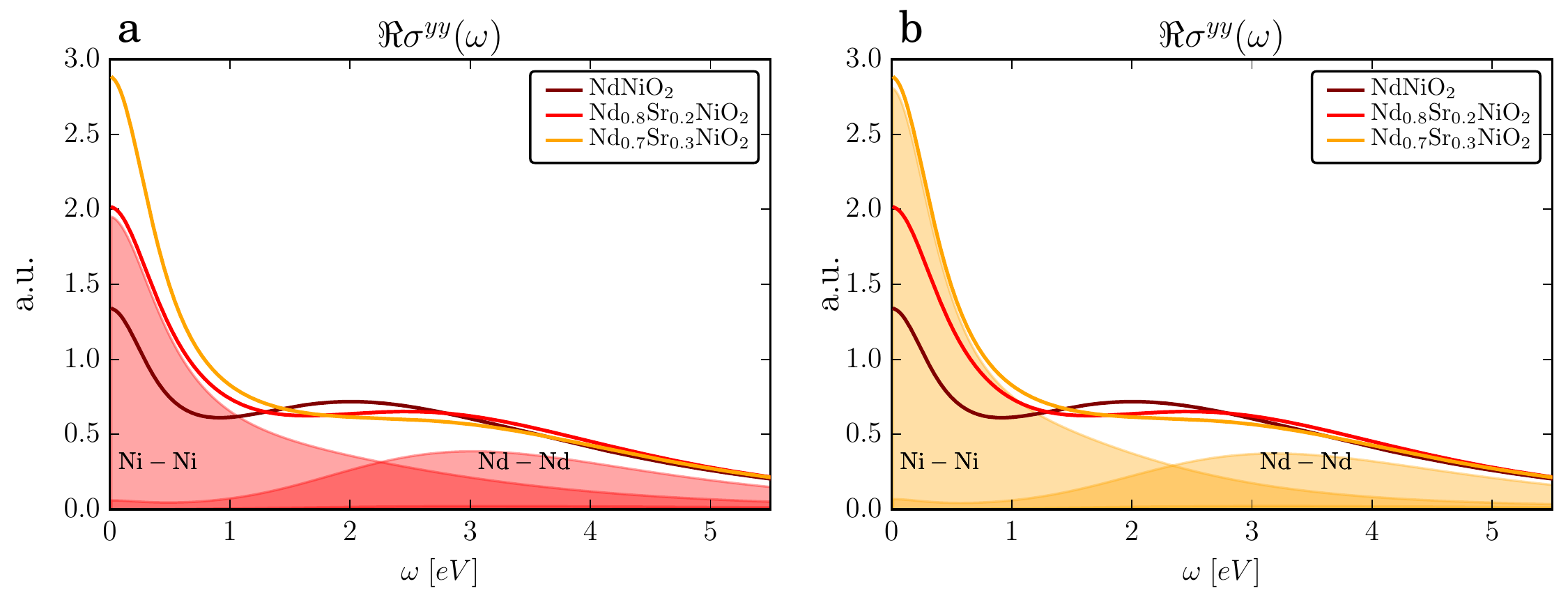}
\par\end{centering}
\caption{Site-resolved optical conductivities for different doping levels.}\label{optcond_suppl}
\end{figure}

%
%
%
%
%
%
%
%
%
%
%
%
%
%
%
\newpage
\begin{center}
\textbf{\large Normal state of Nd$_{1-x}$Sr$_x$NiO$_2$ from self-consistent GW+EDMFT\\ Supplemental Material}
\vspace*{8mm}
\end{center}
\noindent
{\bf Electron density of Nd$_{1-\delta}$Sr$_{\delta}$NiO$_2$.}
In Fig.~\ref{density} we report the orbital resolved density expectation values for the Ni and Nd sites included in the model, and the three doping levels considered. The Nd histograms show how the dopant hole mainly goes to the 
5$d_{z^2}$ orbital while the overall density on the Nickel sites remains approximately pinned to 8.3 electrons.
Despite the fixed electron density, charge within the Ni orbitals is reshuffled with doping in a non-trivial way that, 
in combination with the modified interaction parameters, results in an increased metallicity.
%
%
%
\\
\\
{\bf Calculation of the current correlators.}
Within linear response theory conductivities are computed from current correlators $\Pi$, 
\begin{align}
\frac{1}{e^{2}}\Re\sigma^{yy}(i\nu_{n})	&=i\frac{\Pi^{yy}(i\nu_{n})-\Pi^{yy}(0)}{i\nu_{n}}, \nonumber \\
\frac{2}{e^{3}}\Re\sigma^{yxz}(i\nu_{n})&=\frac{\Pi^{yxz}(i\nu_{n})}{i\nu_{n}}.\end{align}
The optical conductivity 
$\sigma^{yy}$ has been computed for several real materials using DFT+DMFT implementations. \cite{Oudovenko2004,Tomczak2009,Wissgott2012,Deng2016}
This is possible thanks to simplifications that occur in the limit of infinite dimensions, in which the DMFT self-consistency equations are derived. 
In this limit the vertex function becomes local and, in systems with inversion symmetry 
$(v_{\mathbf{k}}^{\mathsf{a}}=\frac{\partial\mathcal{H}\left(\mathbf{k}\right)}{\partial k_{\mathsf{a}}}=-v_{-\mathbf{k}}^{\mathsf{a}})$, its contribution vanishes.
This leads to the evaluation of bubble-like diagrams, sketched in Fig.~\ref{diagram_optc}, involving the Green's functions and band velocities.
To obtain $\Pi^{yy}(\omega)$ we first Fourier transform the correlator evaluated on the imaginary time axis, 
$\Pi_{\alpha\alpha\beta\beta}^{yy}(i\nu_{n})=\mathcal{F}\left\{ \sum_{\mathbf{k}}v_{\alpha\mathbf{k}}^{y}G_{\alpha\beta}(\mathbf{k},\tau)
v_{\beta\mathbf{k}}^{y}G_{\beta\alpha}(\mathbf{k},-\tau)\right\} $, and then perform the analytical continuation to the real frequency axis using the maximum entropy method.\cite{Jarrell1996} 
In Fig.~\ref{optcond_suppl} we report the site-resolved conductivities for the optimally doped and overdoped systems. 
The inter-site Ni-Nd contributions turn out to be negligible compared to the intra-site contributions. It is also interesting 
to notice how the 
feature located at $\sim$1.8eV in the undoped system vanishes in favour of a more 
prominent Drude peak. This is a clear indication that the Nickel becomes less correlated and more metallic if the doping 
is increased and, as noted in the main text, represents a fingerprint of the multi-orbital nature of the system.
\begin{figure}
\begin{centering}
\includegraphics[scale=0.9]{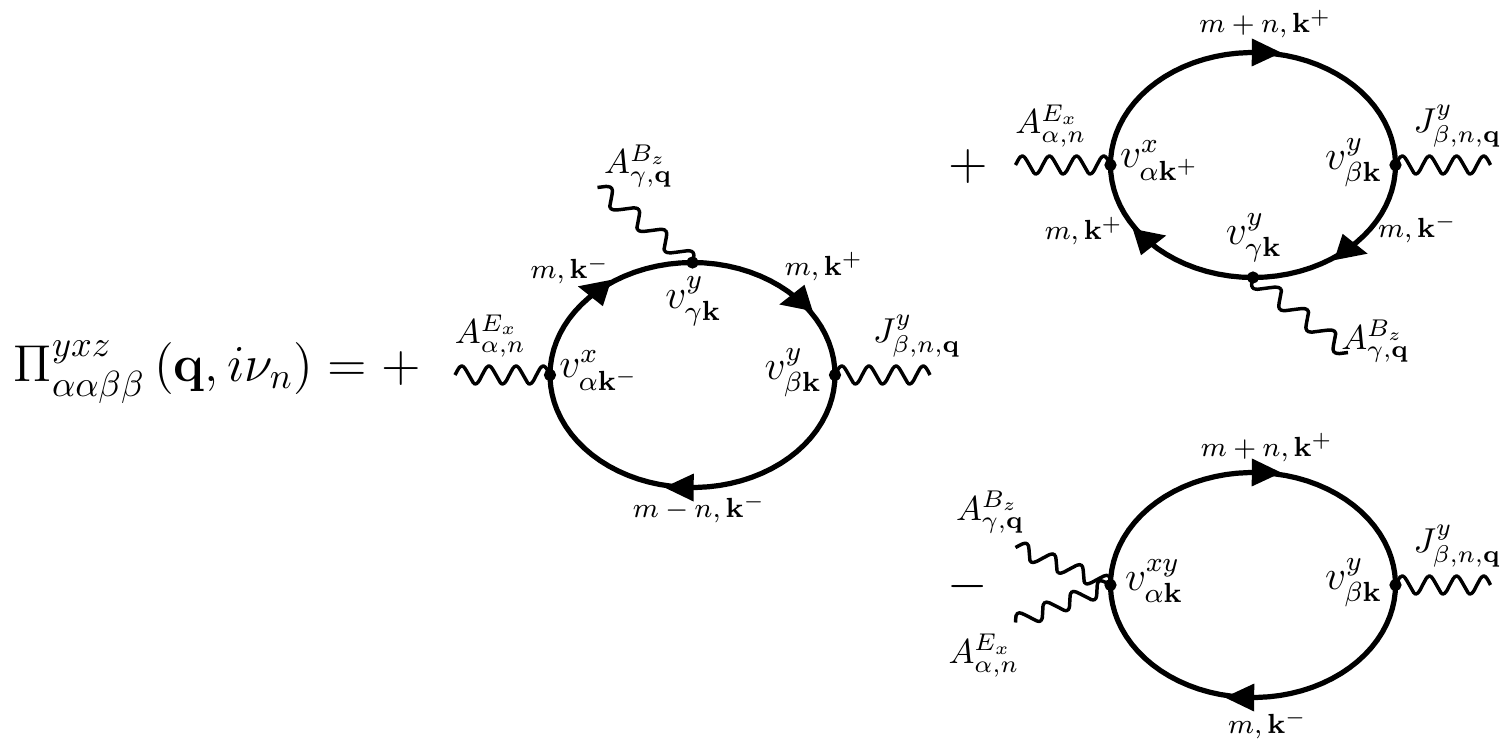}
\par\end{centering}
\caption{Third order diagrams contributing to the Hall conductivity for electric and magnetic fields oriented in the $x$ and $z$ 
directions respectively, and a Hall current in the $y$ direction.}\label{diagram_Hall}
\end{figure}
\\
\\
While the correlator which defines the optical conductivity describes a second-order process, the Hall conductivity 
$\Re\sigma^{yxz}\left(\omega\right)$ stems from a third order process. Strictly following the notation of Ref.~\onlinecite{Voruganti1992}, 
we report in Fig.~\ref{diagram_Hall} the non-vanishing diagrams linear both in the electric $A_{n}^{E_x}$ and magnetic $A_{\mathbf{q}}^{B_z}$ 
field components of the total vector potential, where $\mathbf{k}^{\pm}\equiv\mathbf{k}\pm\frac{\mathbf{q}}{2}$. 
The sum of the reported diagrams vanishes at $\mathbf{q}=0$ and an expansion up to the linear order in $\mathbf{q}$ of both the vertex 
current $v_{\alpha,\mathbf{k}}$ and the Green's function $G\left(\mathbf{k}\pm\frac{\mathbf{q}}{2},i\omega_{n}\right)$ is needed. 
In standard DMFT calculations the locality of the self-energy implies that $G\left(\mathbf{k}\pm\frac{\mathbf{q}}{2},i\omega_{n}\right)=
G\left(\mathbf{k},i\omega_{n}\right)\mp\frac{\mathbf{q}}{2}v_{\mathbf{k}}G^{2}\left(\mathbf{k},i\omega_{n}\right)$. 
This allows for the compact notation used in Refs.~\onlinecite{Lange1998,Haule2003}, but is not a viable option in our case, 
where the $\mathbf{k}$-dependence of the Green's function does not come solely from the dispersion.
Following the procedure outlined for the optical conductivity, we start with the evaluation of the three correlators in imaginary time at small wavevector:
\begin{align}
\Pi_{\alpha\alpha\beta\beta}^{yxz}\left(\delta\mathbf{q},\tau\right)  &	=  \sum_{\mathbf{k}\gamma}\int_{0}^{\tau}d\tau_{1}v_{\alpha\mathbf{k}^{-}}^{x}G_{\alpha\gamma}\left(\mathbf{k}^{-},\tau_{1}\right)v_{\gamma\mathbf{k}}^{y}G_{\gamma\beta}\left(\mathbf{k}^{+},\tau-\tau_{1}\right)v_{\beta\mathbf{k}}^{y}G_{\beta\alpha}\left(\mathbf{k}^{-},-\tau\right) \nonumber \\
	&+\sum_{\mathbf{k}\gamma}\int_{\tau}^{0}d\tau_{1}v_{\alpha\mathbf{k}^{+}}^{x}G_{\alpha\beta}\left(\mathbf{k}^{+},\tau\right)v_{\beta\mathbf{k}}^{y}G_{\beta\gamma}\left(\mathbf{k}^{+},\tau_{1}-\tau\right)v_{\gamma\mathbf{k}}^{y}G_{\gamma\alpha}\left(\mathbf{k}^{-},-\tau_{1}\right) \nonumber \\
	&-\sum_{\mathbf{k}}v_{\alpha\mathbf{k}}^{xy}G_{\alpha\beta}\left(\mathbf{k}^{+},\tau\right)v_{\beta\mathbf{k}}^{y}G_{\beta\gamma}\left(\mathbf{k}^{-},-\tau\right),
\end{align}
where we sum over all the possible orbitals affected by the insertion at $\tau_1$. Then we numerically differentiate this correlator 
with respect to $\delta \mathbf{q}$. The results for the undoped and optimally doped compound are reported in Fig.~\ref{PiJJJ}.
\begin{figure}
\begin{centering}
\includegraphics{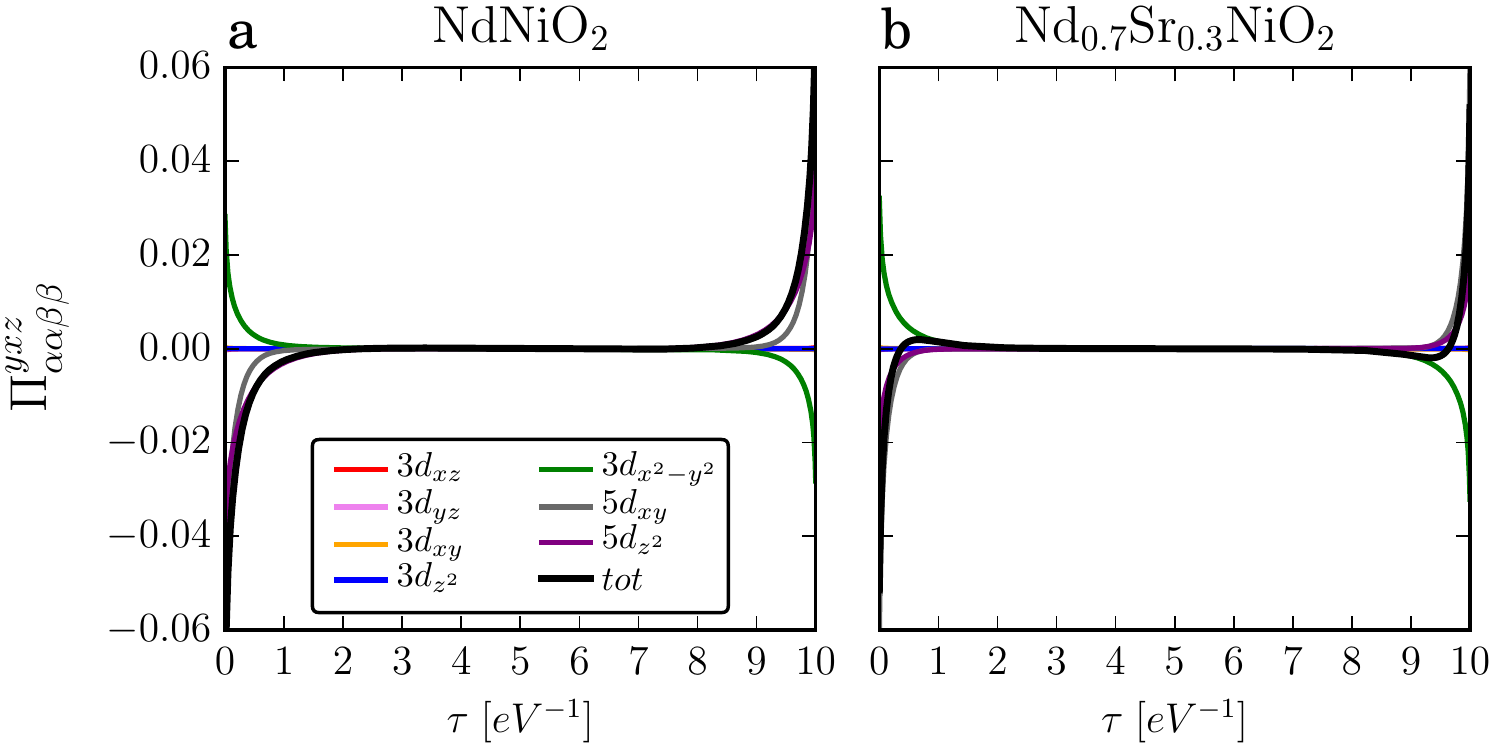}
\par\end{centering}
\caption{Third order current correlators on the imaginary time axis. Only Ni-3$d_{x^2-y^2}$, Nd-$5d_{xy}$ and Nd-$5d_{z^2}$ yield relevant contributions.}\label{PiJJJ}
\end{figure}
We then Fourier transform to Matsubara frequency and perform the analytic continuation to real frequency via the Pad\'e algorithm.\cite{Vidberg1977} 
We checked that the asymptotic behaviour of both the real and imaginary parts of the resulting Hall conductivity are compatible 
with the results derived in Ref.~\onlinecite{Lange1998}. Since analytic continuation algorithms might produce artefacts, 
to obtain $\Re\sigma^{yxz}\left(0\right)$, we assume that $\Pi_{\alpha\alpha\beta\beta}^{yxz}(i\nu_n)$ is purely imaginary and vanishes at $\nu_n=0$ so that
\begin{align}
\frac{2}{e^{3}}\Re\sigma^{yxz}\left(0\right)=\left.\frac{d\Im\Pi\left(\nu_{n}\right)}{d\nu_{n}}\right|_{0},
\end{align}
which allows for a more precise estimate of the Hall coefficient, as plotted in Fig.~5d of the main text.
%
%
%
\newpage
\noindent
{\bf Doping dependent effective local interactions.} We report in the same format as in the main text 
the effective screened interactions produced by the self-consistent solution of the two impurity models.
A relevant observation is that the $3d_{z^2}$-$3d_{x^2-y^2}$ inter-orbital interaction 
increases more strongly as a function of doping than the intra-orbital $3d_{x^2-y^2}$ interaction, 
while the corresponding Hund coupling parameter is reduced. Both effects contribute to an effectively 
weaker correlation strength and an enhanced metallicity in the doped state.
\begin{align}
\begin{array}{cccc}
\hat{\mathcal{I}}_{\mathrm{Ni}}^{\delta=0.0}\left(0\right) & =\begin{array}{c}
\begin{array}{c|ccccc}
 & d_{xz} & d_{yz} & d_{xy} & d_{z^{2}} & d_{x^{2}-y^{2}}\\
\hline d_{xz} & 4.954 & 3.504 & 3.394 & 3.815 & 3.084\\  \cline{2-2}
d_{yz} & \multicolumn{1}{c|}{0.696} & 4.954 & 3.394 & 3.815 & 3.084\\  \cline{3-3}
d_{xy} & 0.689 &  \multicolumn{1}{c|}{0.689} & 4.741 & 3.075 & 3.545\\  \cline{4-4}
d_{z^{2}} & 0.501 & 0.501 & \multicolumn{1}{c|}{0.710} & 4.850 & 2.822\\  \cline{5-5}
d_{x^{2}-y^{2}} & 0.526 & 0.526 & 0.350 & \multicolumn{1}{c|}{0.571} & 3.981
\end{array}
\end{array} &  & \hat{\mathcal{I}}_{\mathrm{Nd}}^{\delta=0.0}\left(0\right)=\begin{array}{c}
\begin{array}{c|cc}
 & d_{xy} & d_{z^{2}}\\
\hline d_{xy} & 1.817 & 1.344\\  \cline{2-2}
d_{z^{2}} &  \multicolumn{1}{c|}{0.351} & 1.682
\end{array}\end{array}\\
\\
\\
\hat{\mathcal{I}}_{\mathrm{Ni}}^{\delta=0.2}\left(0\right) & =\begin{array}{c}
\begin{array}{c|ccccc}
 & d_{xz} & d_{yz} & d_{xy} & d_{z^{2}} & d_{x^{2}-y^{2}}\\
\hline d_{xz} & 5.242 & 3.806 & 3.622 & 4.179 & 3.289\\ \cline{2-2}
d_{yz} & \multicolumn{1}{c|}{0.685} & 5.242 & 3.622 & 4.179 & 3.289\\ \cline{3-3}
d_{xy} & 0.673 & \multicolumn{1}{c|}{0.673} & 4.854 & 3.346 & 3.665\\ \cline{4-4}
d_{z^{2}} & 0.495 & 0.495 & \multicolumn{1}{c|}{0.704} & 5.316 & 3.091\\ \cline{5-5}
d_{x^{2}-y^{2}} & 0.511 & 0.511 & 0.340 & \multicolumn{1}{c|}{0.559} & 4.072
\end{array}\end{array} &  & \hat{\mathcal{I}}_{\mathrm{Nd}}^{\delta=0.2}\left(0\right)=\begin{array}{c}
\begin{array}{c|cc}
 & d_{xy} & d_{z^{2}}\\
\hline d_{xy} & 1.787 & 1.396\\  \cline{2-2}
d_{z^{2}} &  \multicolumn{1}{c|}{0.360} & 1.748
\end{array}\end{array}\\
\\
\\
\hat{\mathcal{I}}_{\mathrm{Ni}}^{\delta=0.3}\left(0\right) & =\begin{array}{c}
\begin{array}{c|ccccc}
 & d_{xz} & d_{yz} & d_{xy} & d_{z^{2}} & d_{x^{2}-y^{2}}\\ 
\hline d_{xz} & 5.306 & 3.879 & 3.629 & 4.316 & 3.300\\ \cline{2-2}
d_{yz} & \multicolumn{1}{c|}{0.663} & 5.306 & 3.629 & 4.316 & 3.300\\ \cline{3-3}
d_{xy} & 0.652 & \multicolumn{1}{c|}{0.652} & 4.769 & 3.417 & 3.613\\ \cline{4-4}
d_{z^{2}} & 0.483 & 0.483 & \multicolumn{1}{c|}{0.687} & 5.536 & 3.148\\ \cline{5-5}
d_{x^{2}-y^{2}} & 0.486 & 0.486 & 0.329 & \multicolumn{1}{c|}{0.545} & 4.014
\end{array}\end{array} &  & \hat{\mathcal{I}}_{\mathrm{Nd}}^{\delta=0.3}\left(0\right)=\begin{array}{c}
\begin{array}{c|cc}
 & d_{xy} & d_{z^{2}}\\
\hline d_{xy} & 1.726 & 1.350\\  \cline{2-2}
d_{z^{2}} &  \multicolumn{1}{c|}{0.360} & 1.677
\end{array}\end{array}
\end{array}
\end{align}
%
%
%
\noindent
{\bf Orbitally resolved spectral functions of Nd$_{1-\delta}$Sr$_{\delta}$NiO$_2$.} In Figs.~\ref{AkwNi_0}-\ref{AkwNd_2} we report the orbitally resolved
dispersions for the seven orbitals considered in our study. All the results are obtained from the $\mathbf{k}$ dependent 
spectral function so as to preserve the orbital character for each band. 
In the LDA case this is directly evaluated, while at the $G^0W^0$ and  $GW$+EDMFT level they are compute by analytical continuation at each $\mathbf{k}$ point.
\begin{figure}[H]
\begin{centering}
\includegraphics[scale=0.7]{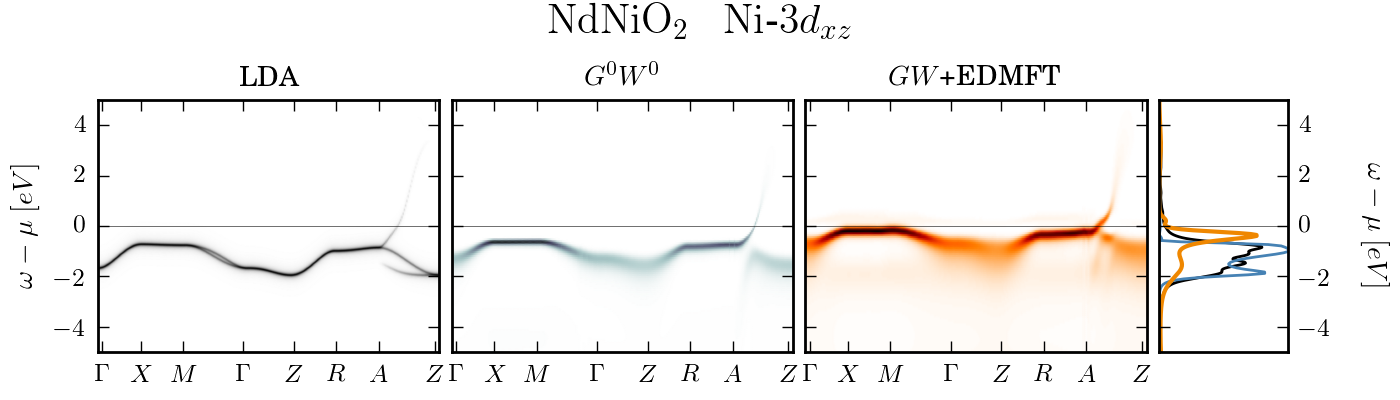}
\par\end{centering}
\begin{centering}
\includegraphics[scale=0.7]{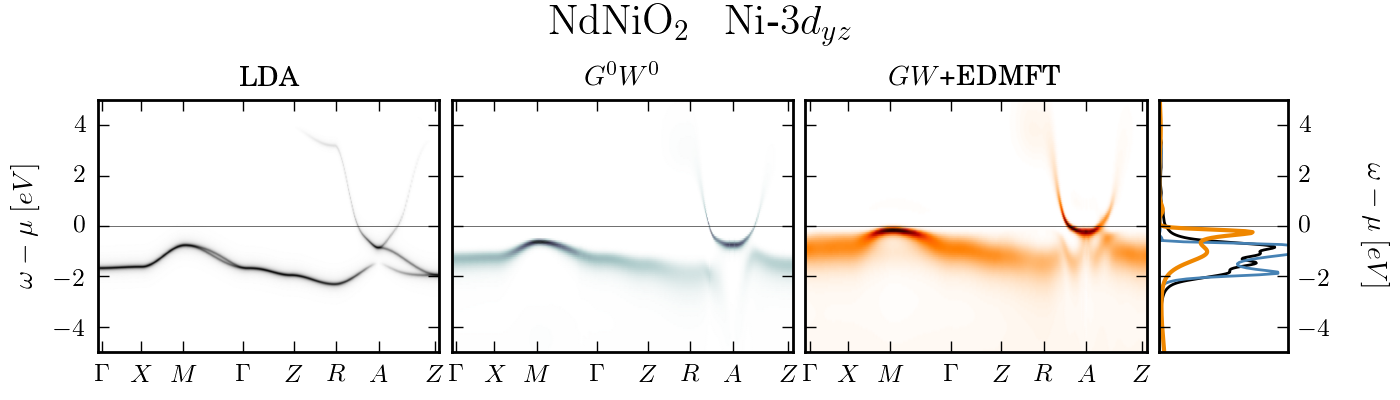}
\par\end{centering}
\begin{centering}
\includegraphics[scale=0.7]{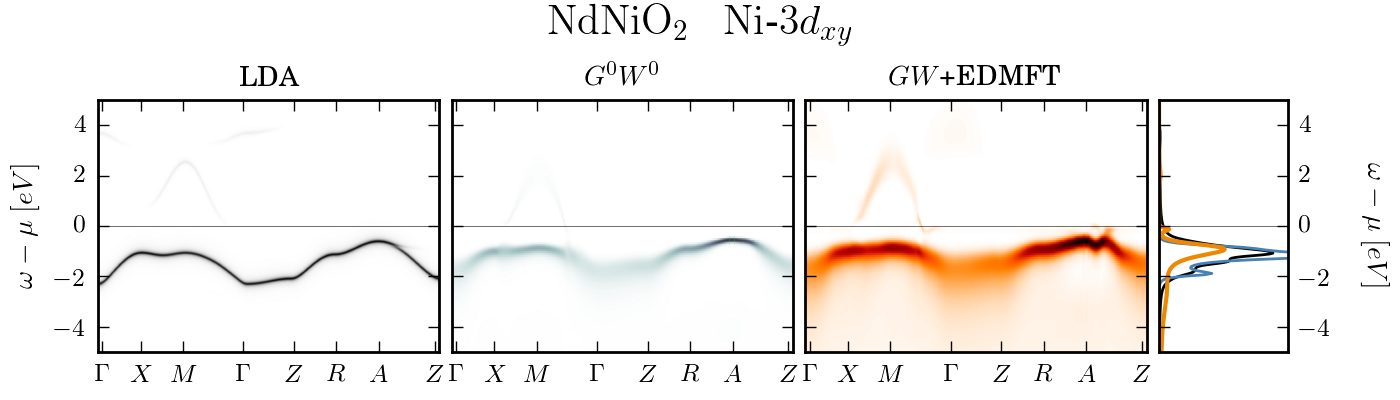}
\par\end{centering}
\begin{centering}
\includegraphics[scale=0.7]{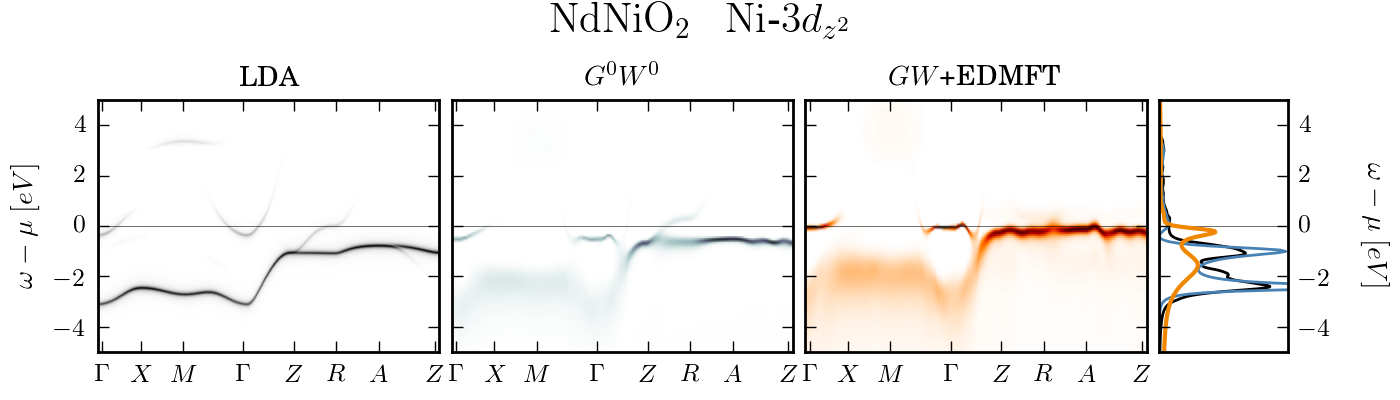}
\par\end{centering}
\begin{centering}
\includegraphics[scale=0.7]{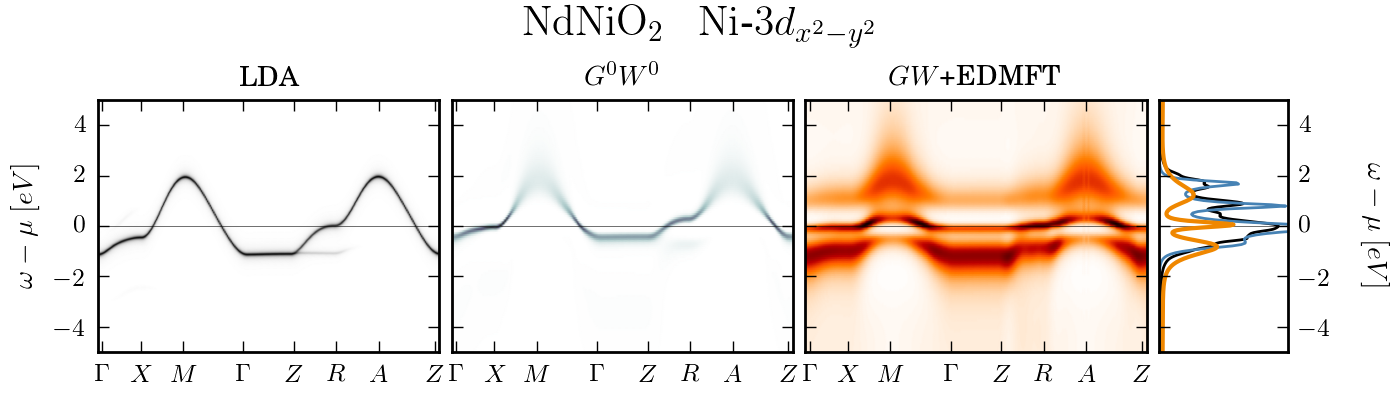}
\par\end{centering}
\caption{Nickel spectral functions for the undoped setup.}\label{AkwNi_0}
\end{figure}
\begin{figure}[H]
\begin{centering}
\includegraphics[scale=0.7]{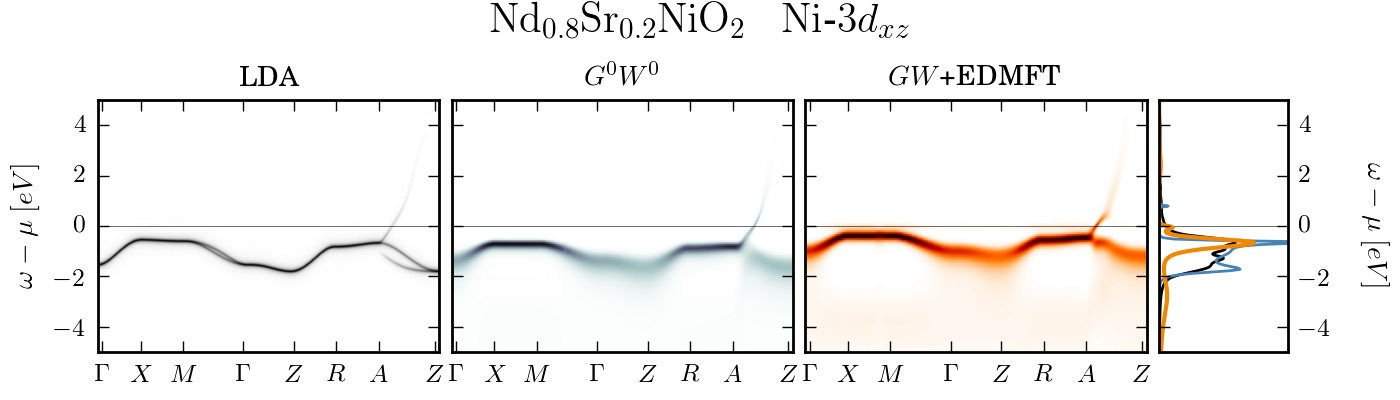}
\par\end{centering}
\begin{centering}
\includegraphics[scale=0.7]{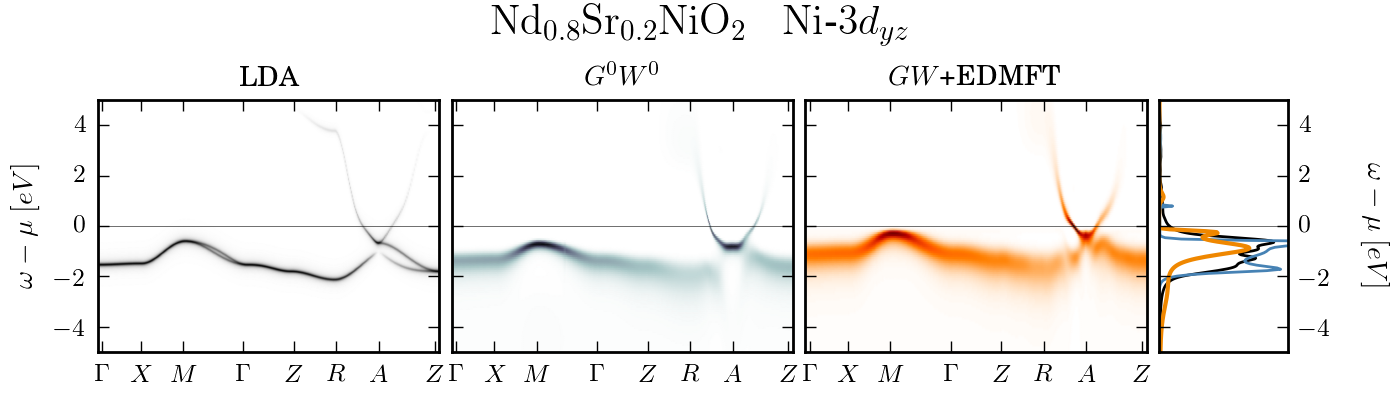}
\par\end{centering}
\begin{centering}
\includegraphics[scale=0.7]{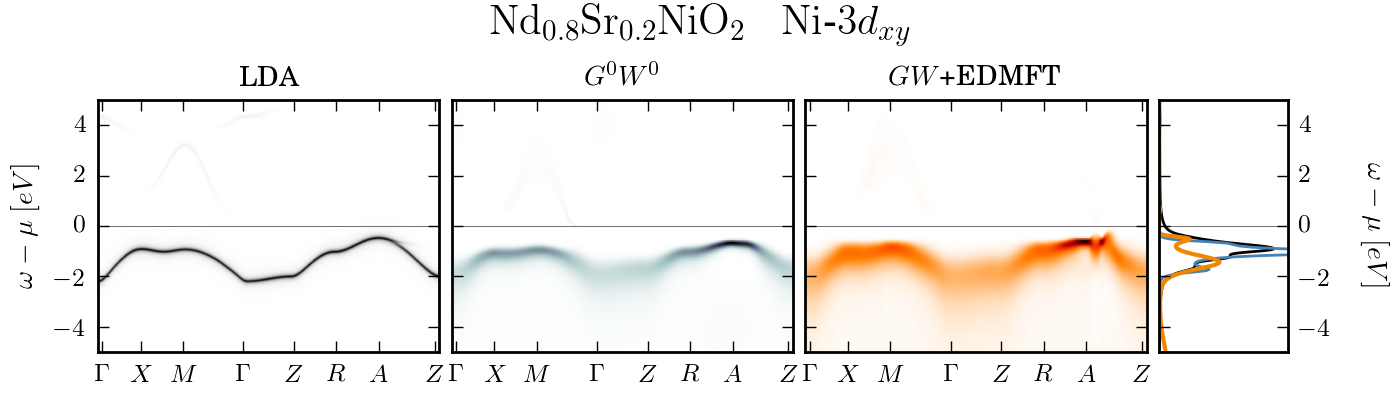}
\par\end{centering}
\begin{centering}
\includegraphics[scale=0.7]{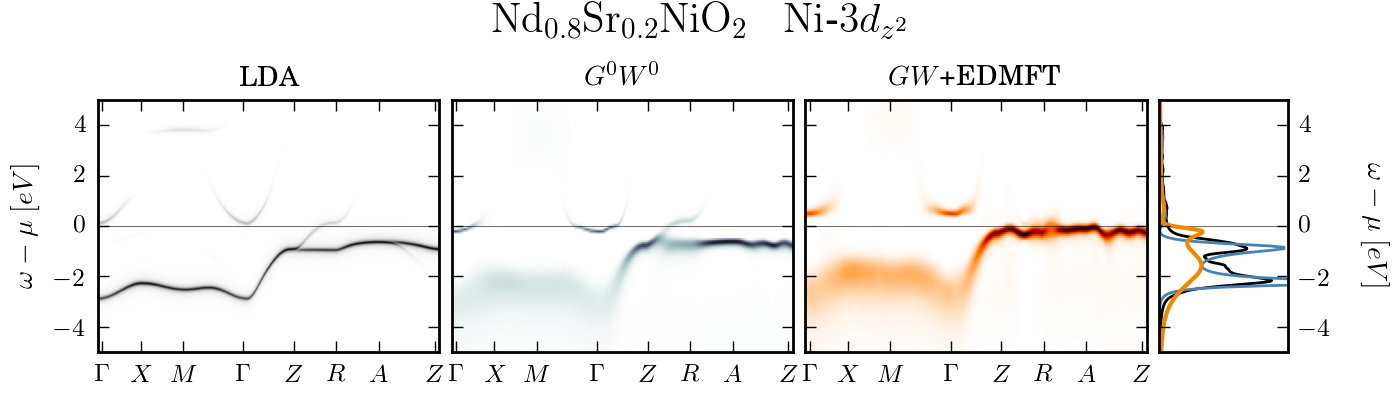}
\par\end{centering}
\begin{centering}
\includegraphics[scale=0.7]{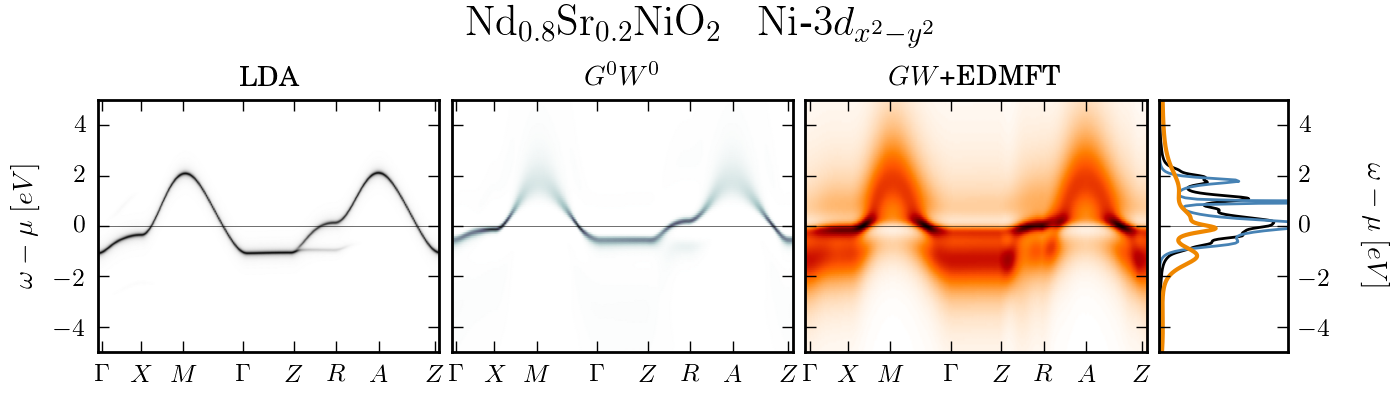}
\par\end{centering}
\caption{Nickel spectral functions for the optimally doped setup.}\label{AkwNi_1}
\end{figure}
\begin{figure}[H]
\begin{centering}
\includegraphics[scale=0.7]{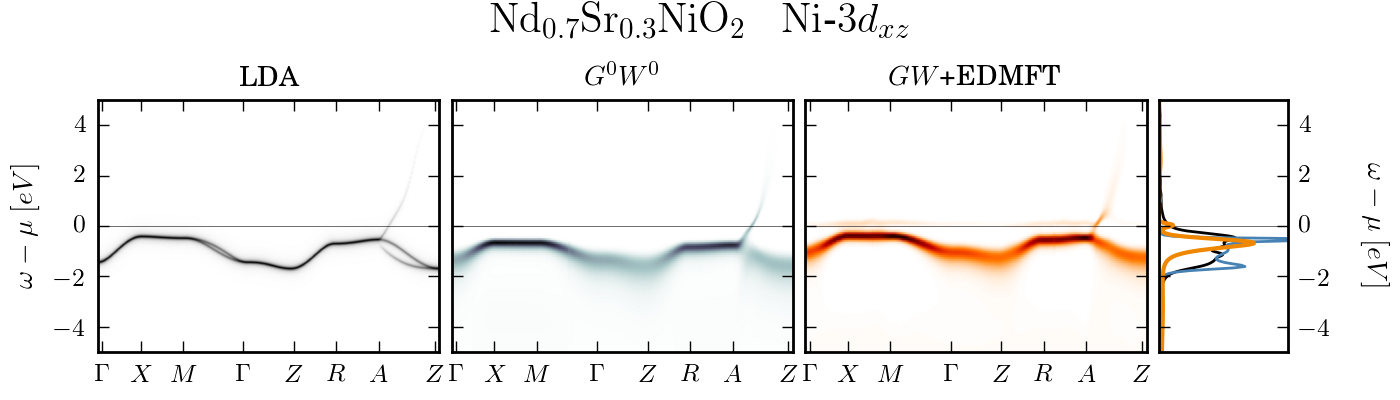}
\par\end{centering}
\begin{centering}
\includegraphics[scale=0.7]{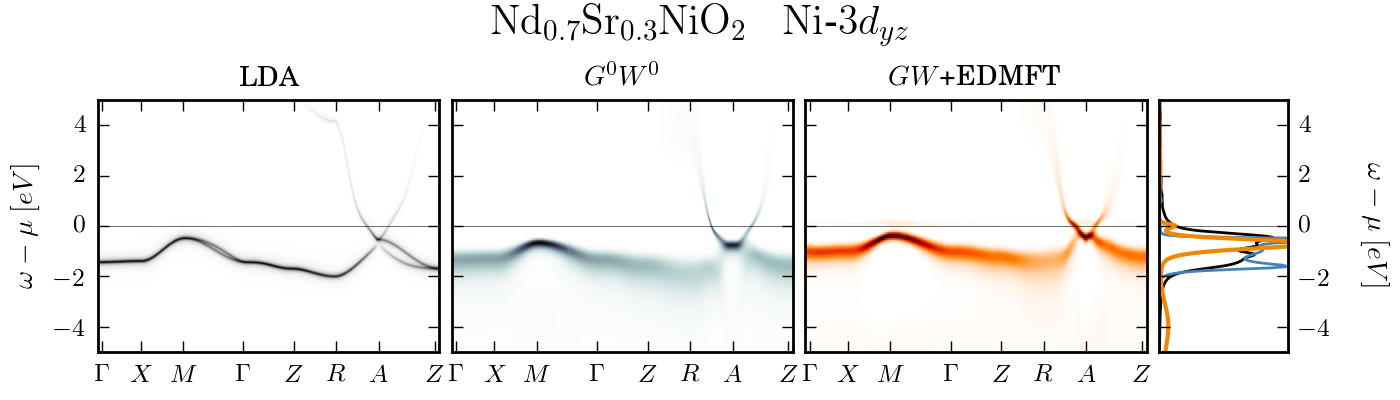}
\par\end{centering}
\begin{centering}
\includegraphics[scale=0.7]{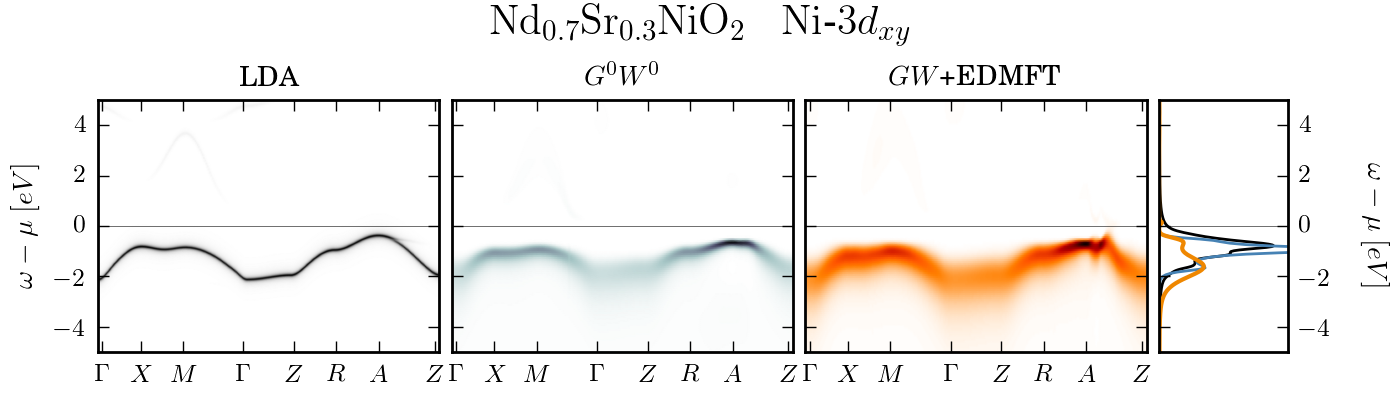}
\par\end{centering}
\begin{centering}
\includegraphics[scale=0.7]{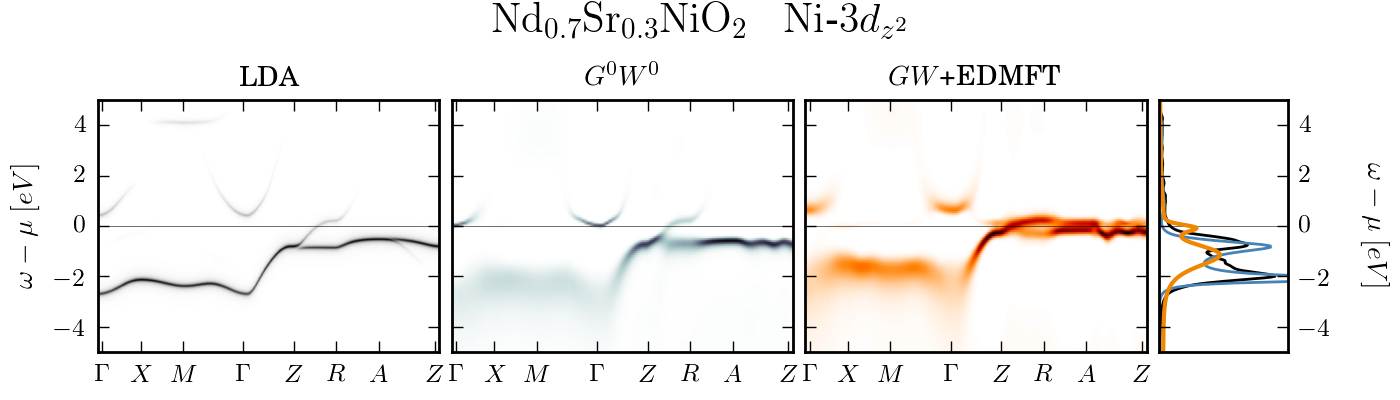}
\par\end{centering}
\begin{centering}
\includegraphics[scale=0.7]{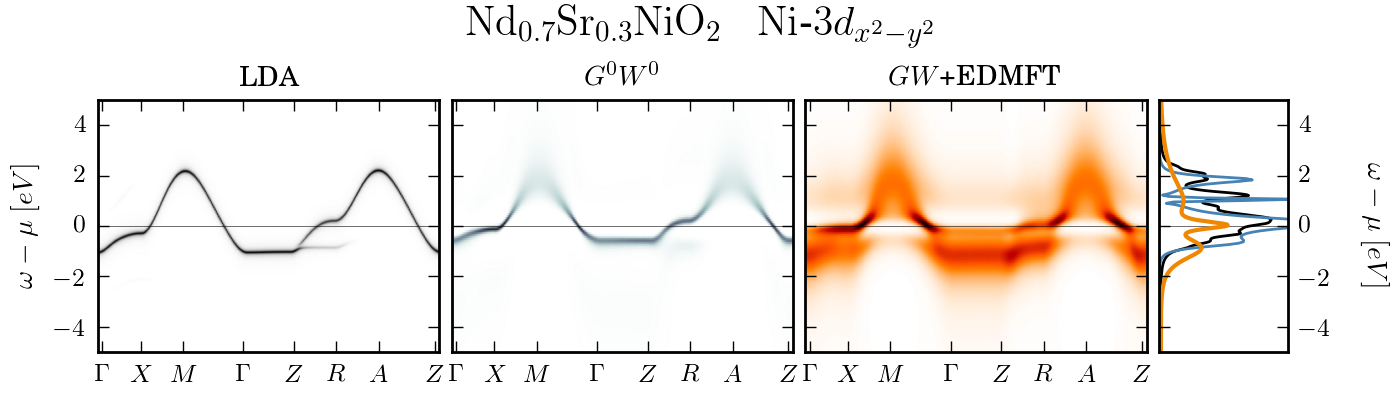}
\par\end{centering}
\caption{Nickel spectral functions for the overdoped doped setup.}\label{AkwNi_2}
\end{figure}
\begin{figure}[H]
\begin{centering}
\includegraphics[scale=0.7]{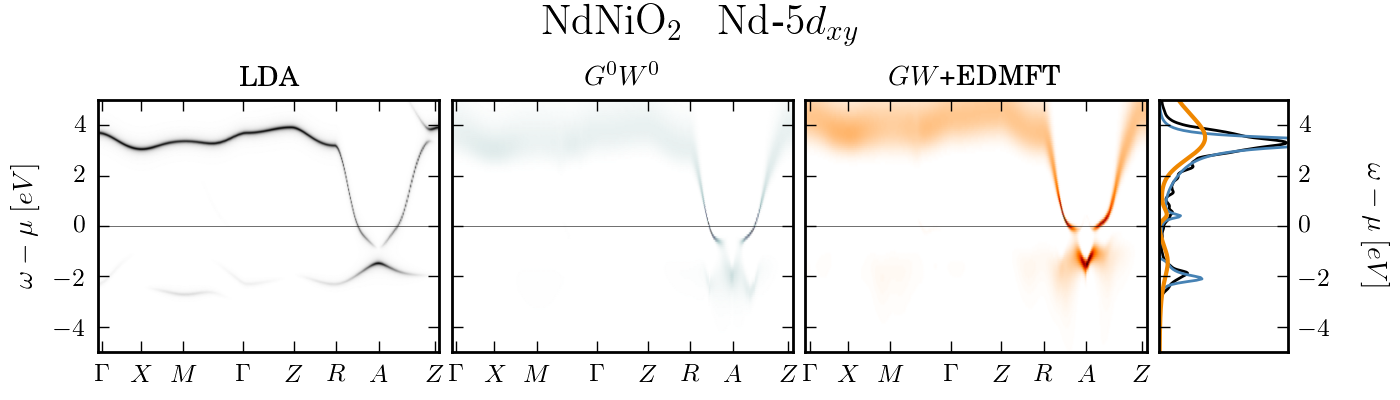}
\par\end{centering}
\begin{centering}
\includegraphics[scale=0.7]{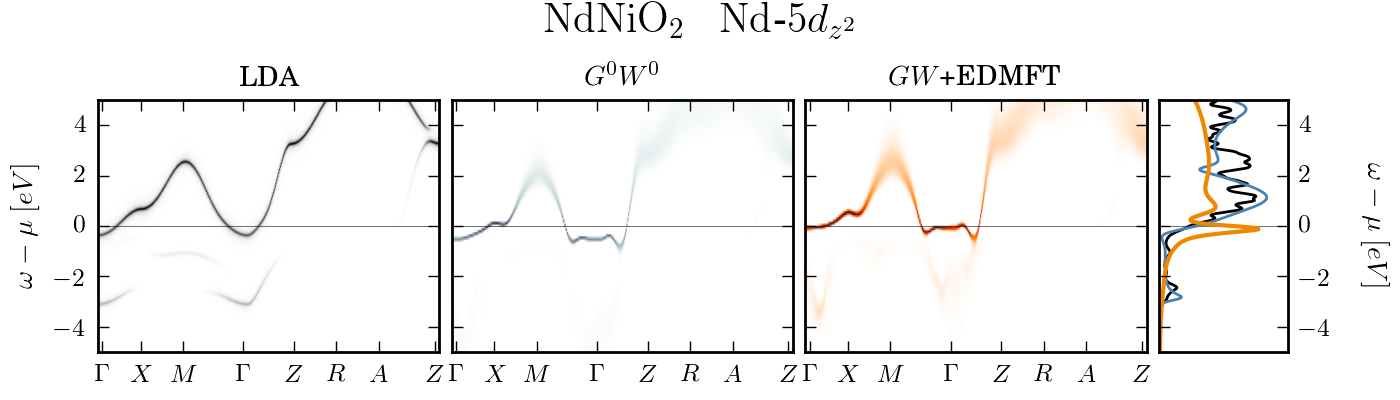}
\par\end{centering}
\caption{Neodymium spectral functions for the undoped setup.}\label{AkwNd_0}
\end{figure}
\begin{figure}[H]
\begin{centering}
\includegraphics[scale=0.7]{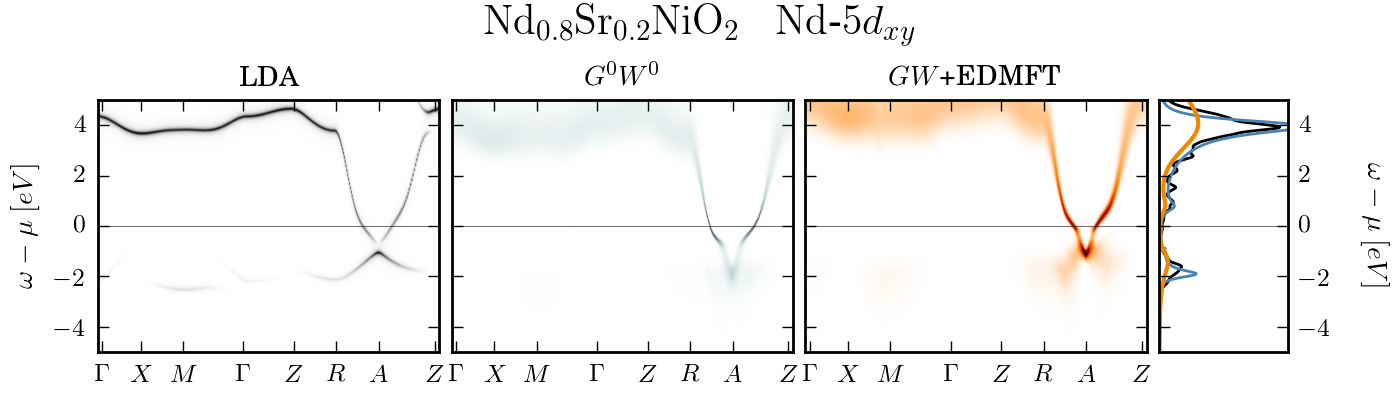}
\par\end{centering}
\begin{centering}
\includegraphics[scale=0.7]{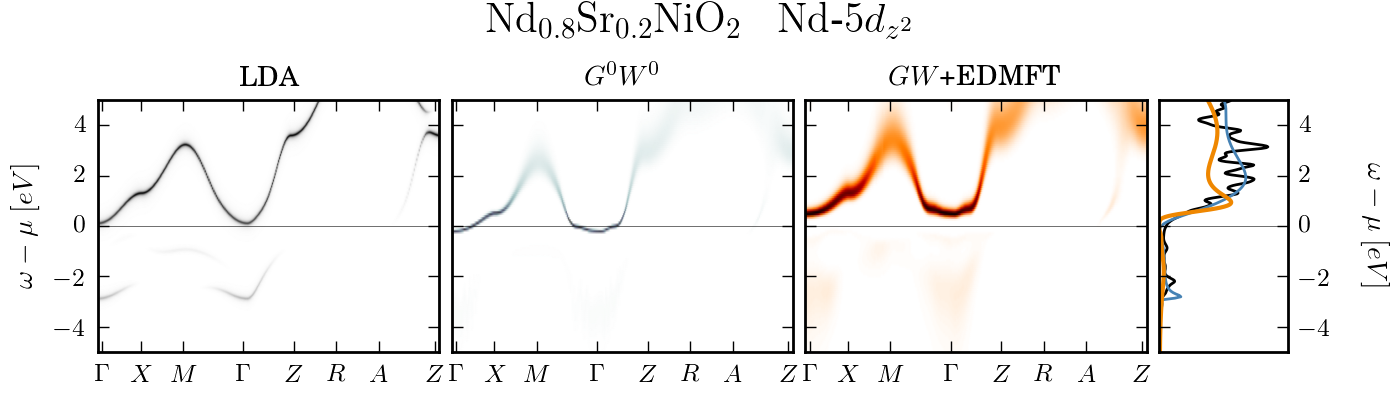}
\par\end{centering}
\caption{Neodymium spectral functions for the optimally doped setup.}\label{AkwNd_1}
\end{figure}
\begin{figure}[H]
\begin{centering}
\includegraphics[scale=0.7]{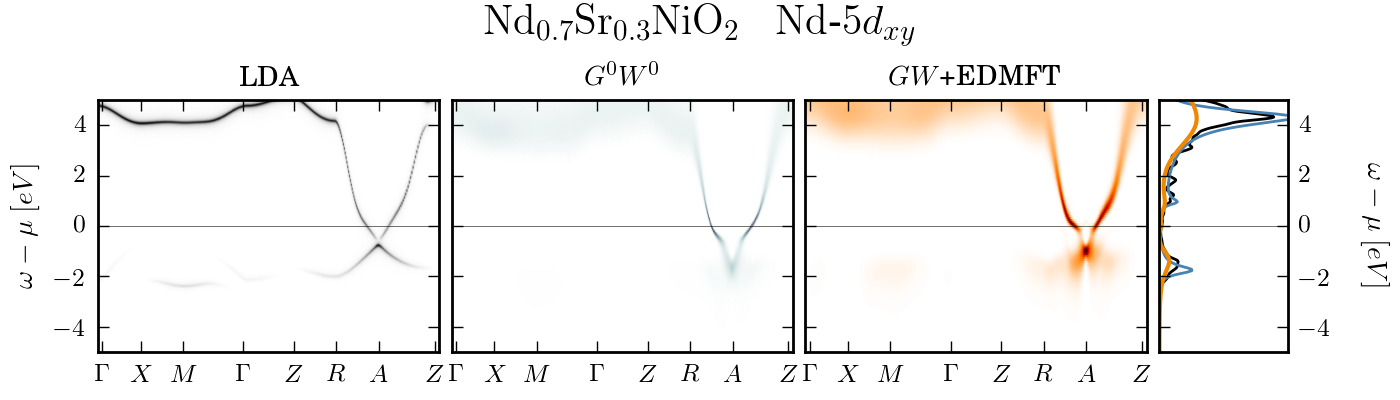}
\par\end{centering}
\begin{centering}
\includegraphics[scale=0.7]{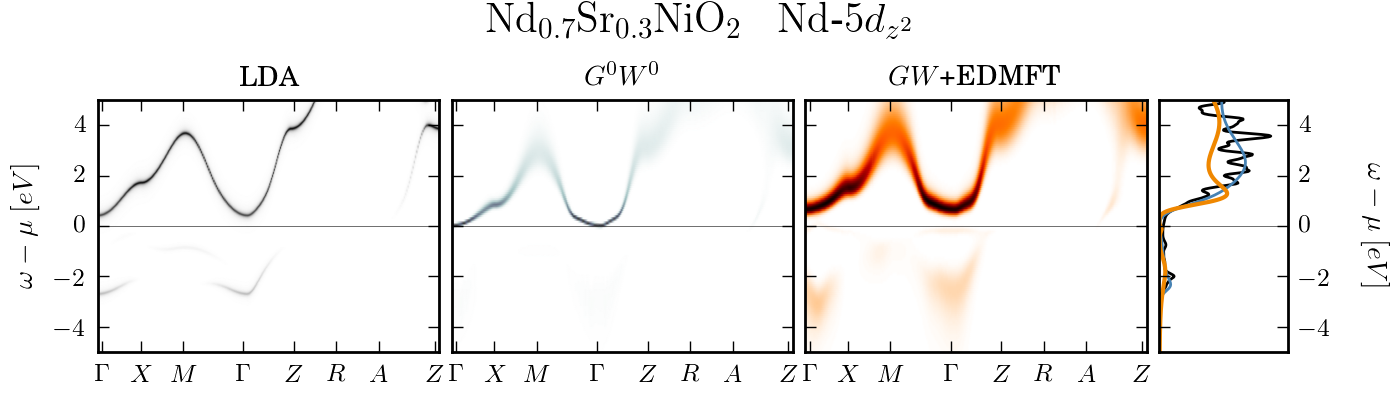}
\par\end{centering}
\caption{Neodymium spectral functions for the overdoped setup.}\label{AkwNd_2}
\end{figure}
%
%
%
\newpage
\noindent
{\bf Interacting Fermi surfaces.} In Fig.~\ref{Fermi_suppl} we report the Fermi surfaces for the Ni-3$d_{z^2}$ and Ni-3$d_{x^2-y^2}$ with the same arrangement 
as in the main text. The color intensities are normalized to the maximum value for each orbital, and the different $k_z$ panels (rows of the plot ) use the same normalization.
We notice that at $k_z=0.5$ the Ni-3$d_{x^2-y^2}$ surface get closer to the one of CaCuO$_2$ reported in Ref.~\onlinecite{Karp2020}.
\begin{figure}[H]
\begin{centering}
\includegraphics[scale=0.8]{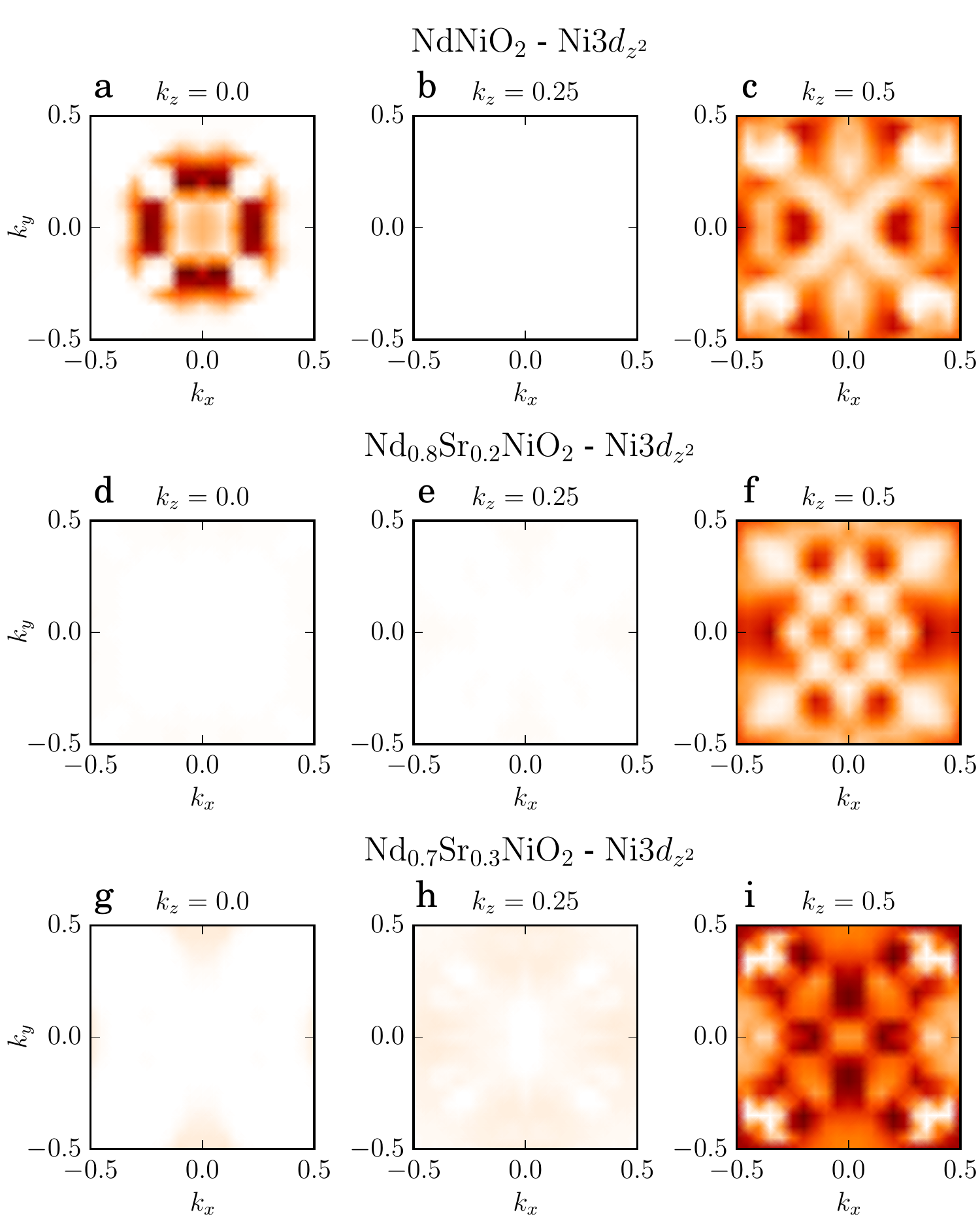}
\par\end{centering}
\caption{Fermi surfaces of Ni-3$d_{z^2}$. For a fixed doping level, we use the same color range for the different $k_z$.
}\label{Fermi_suppl}
\end{figure}
\begin{figure}[H]
\begin{centering}
\includegraphics[scale=0.8]{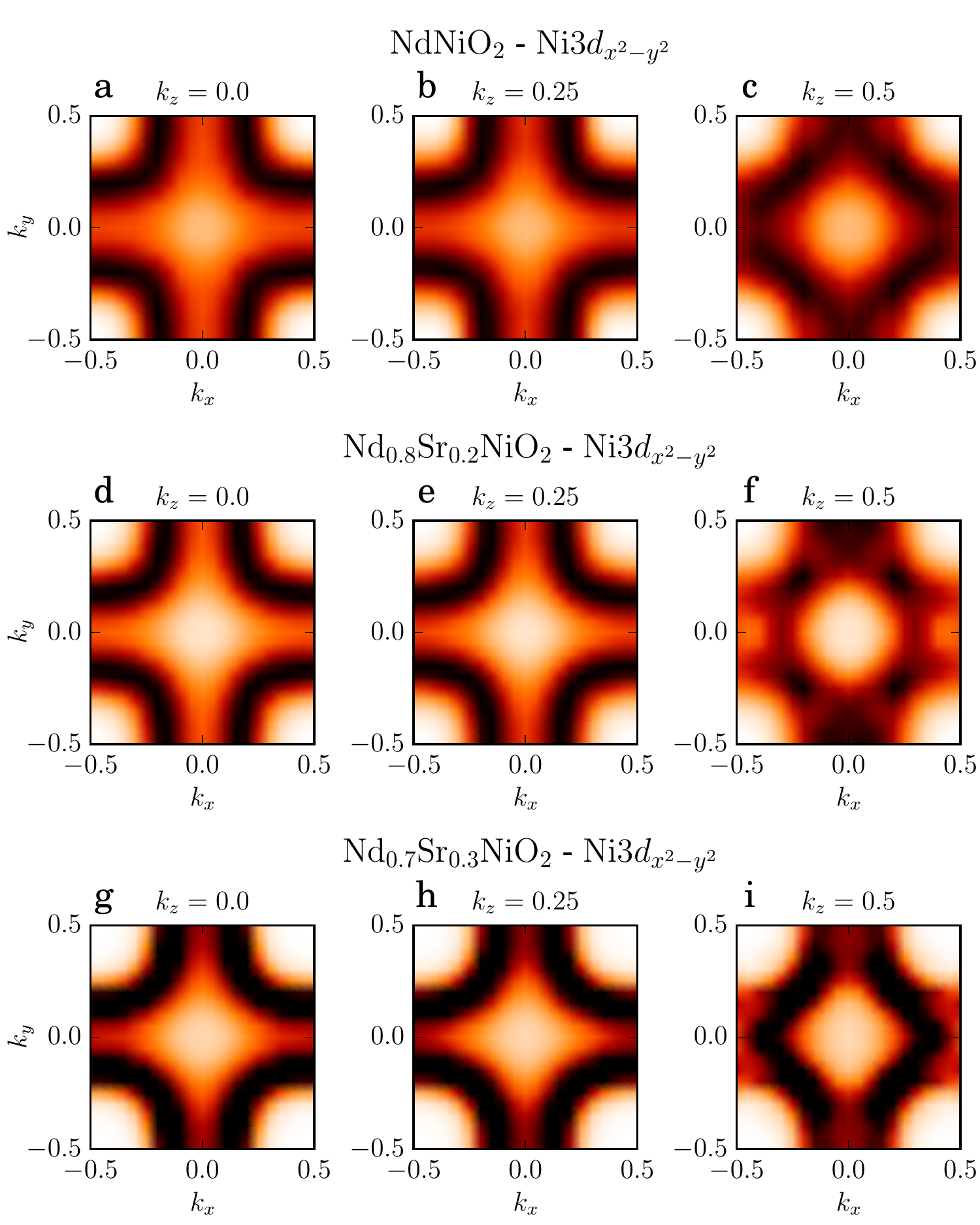}
\par\end{centering}
\caption{Fermi surfaces of Ni-3$d_{x^2-y^2}$. For a fixed doping level, we use the same color range for the different $k_z$.
}\label{Fermi_suppl}
\end{figure}

\begin{thebibliography}{99}
\bibitem{note1}This is not in general valid for realistic dispersions but, considering that the $GW$ vertex function is a delta function and the EDMFT contribution is local, we consider this approximation accurate enough.
\end{thebibliography}
\end{document}